\documentclass{aa}
\usepackage[varg]{txfonts}
\usepackage{amsmath}
\usepackage{siunitx}
\usepackage{xfrac}

\usepackage{natbib}
\usepackage{color}
\usepackage{graphicx}
\usepackage{booktabs}
\usepackage{csquotes}
\makeatletter
  \renewcommand*\aa@pageof{, page \thepage{} of \pageref*{LastPage}}
\makeatother
\usepackage{hyperref}
\hypersetup{pdfpagemode = {UseNone},
            pdftitle = {Migration of planets in circumbinary discs},
            pdfcreator = {\LaTeX},
            pdfproducer = {pdfeTeX-0.\the\pdftexversion\pdftexrevision},
            pdfauthor = {Daniel Thun, Wilhelm Kley},
            pdfsubject = {},
            pdfview = {FitH},
            pdfstartview = {FitH},
            colorlinks = {true},
            linkcolor = [rgb]{0,0.35,0.7},
            citecolor = [rgb]{0,0.35,0.7},
            filecolor = [rgb]{0.61,0,0},
            urlcolor = [rgb]{0.61,0,0},
           }

\usepackage{bm}
\renewcommand{\vec}[1]{\bm{\mathrm{#1}}}

\usepackage{balance}

\newcommand*\dd[1]{\,\mathrm{d}#1}

\newcommand{\ep}{e_\mathrm{p}}
\newcommand{\ap}{a_\mathrm{p}}

\begin{document}

\title{Migration of planets in circumbinary discs}

\author{Daniel Thun \inst{1} \and
        Wilhelm Kley \inst{1}}

\institute{
Institut für Astronomie und Astrophysik, Universität Tübingen,
Auf der Morgenstelle 10, D-72076 Tübingen, Germany\\
\email{\{daniel.thun@, wilhelm.kley@\}uni-tuebingen.de}\\
}

\date{}

\abstract
{}
{
The discovery of planets in close orbits around binary stars raises questions
about their formation. It is believed that these planets formed in the outer
regions of the disc and then migrated through planet-disc interaction to their
current location.  Considering five different systems (Kepler-16, -34, -35, -38
and -413) we model planet migration through the disc, with special focus on the
final orbital elements of the planets.  We investigate how the final orbital
parameters are influenced by the disc and planet masses.
}
{
Using two-dimensional, locally isothermal, viscous hydrodynamical simulations we
first model the disc dynamics for all five systems, followed by a study of the
migration properties of embedded planets with different masses.  To strengthen
our results, we apply two grid-based hydrodynamical codes using different
numerics (\textsc{Pluto} and \textsc{Fargo3D}).
}
{
For all systems, we find that the discs become eccentric and precess slowly.  We
confirm the bifurcation feature in the precession period -- gap-size diagram for
different binary mass ratios.  The Kepler-16, -35, -38 and -413 systems lie on
the lower branch and Kepler-34 on the upper one. For systems with small binary
eccentricity, we find a new non-monotonic, loop-like feature. 

In all systems, the planets migrate to the inner edge of the disc cavity.
Depending on the planet-disc mass ratio, we observe one of two different
regimes.  Massive planets can significantly alter the disc structure by
compressing and circularising the inner cavity and they remain on nearly
circular orbits.  Lower-mass planets are strongly influenced by the disc, their
eccentricity is excited to high values, and their orbits are aligned with the
inner disc in a state of apsidal corotation.
 
In our simulations, the final locations of the planets are typically too large
with respect to the observations because of the large inner gaps of the discs.
The migrating planets in the most eccentric discs (around Kepler-34 and -413)
show the largest final eccentricity in agreement with the observations.
}
{}


\maketitle

\section{Introduction}\label{sec:intro}
Following the first detection of a circumbinary planet with the \emph{Kepler
space telescope}, namely Kepler-16b, eight more binary star systems with a
planet on a P-type orbit have been discovered\footnote{Kepler-16
\citep{2011Sci...333.1602D}, Kepler-34 and -35 \citep{2012Natur.481..475W},
Kepler-38 \citep{2012ApJ...758...87O}, Kepler-47 \citep{2012Sci...337.1511O},
Kepler-64 \citep{2013ApJ...768..127S}, Kepler-413 \citep{2014ApJ...784...14K},
Kepler-453 \citep{2015ApJ...809...26W}, and Kepler-1647 \citep{Kostov:2016}}.
All these systems show striking similarities. They are all very flat, meaning
that the binary and the planet orbit are in the same plane, suggesting that
these planets formed in a circumbinary disc aligned with the orbital plane of
the central binary.  Furthermore, in all systems, the innermost planet (so far
only Kepler-47 is known to have more than one planet) is close to the calculated
stability limit \citep{1986A&A...167..379D,1999AJ....117..621H}.  This leads us
to question  where in the disc these planets formed. Two scenarios are possible:
An in situ formation at the observed location, or a formation further outside
the disc followed by radial migration of the planet through the disc to the
observed location. Strong gravitational interaction between the binary and the
disc, which leads to the excitation of spiral waves in the disc, makes an in
situ formation unlikely \citep{2007A&A...472..993P, 2012ApJ...754L..16P,
2012ApJ...761L...7M, 2015ApJ...808...58S}, because for orbits close to the
binary, destructive collisions between planetesimals are expected. The different
alignment of their periapses lead to high relative impact velocities
\citep{2007MNRAS.380.1119S, 2008ApJ...681.1599M}.

Therefore, it is believed that these circumbinary planets on orbits close to the
stability limit were formed in the outer disc where the binary has less
influence on the formation process. In order to reach their observed position,
the planets then migrated through the disc to their present location either
through type-I or type-II migration \citep{1986ApJ...309..846L,
1997Icar..126..261W, 2000MNRAS.318...18N, 2002ApJ...565.1257T}. This leads to
the question of how this migration progress can be stopped at the right orbit.

Early work by \citet{1994ApJ...421..651A} showed that through tidal interaction
between the binary and the disc an eccentric inner gap in the disc forms. The
structure of the gap, such as eccentricity and size, depends on disc parameters
(viscosity, pressure) as well as on binary parameters (eccentricity, mass
ratio).  This was confirmed by various studies, which all showed an eccentric
inner gap that slowly precesses in a prograde manner around the binary
\citep{2008A&A...483..633P, 2013A&A...556A.134P, 2014A&A...564A..72K,
2015A&A...581A..20K, 2015A&A...582A...5L, 2017MNRAS.466.1170M,
2017MNRAS.465.4735M, 2017A&A...604A.102T}.

In several of these mentioned works, not only was the structure of the
circumbinary disc investigated but also the migration of the planets. These
simulations showed that the inner cavity of the disc constitutes a barrier for
the migrating planet.  The sudden drop of density produces a large positive
corotation torque balancing the negative Lindblad torques responsible for the
inward migration \citep{2006ApJ...642..478M}.

\citet{2013A&A...556A.134P} modelled planets around the systems Kepler-16,
Kepler-34, and Kepler-35. In their numerous simulations, they studied the impact
of the discs' pressure and viscosity onto the migration process of the planets.
They found that the structure (size and eccentricity) of the inner cavity
depends strongly on disc parameters and therefore pressure and viscosity in the
disc have a notable influence on the final orbital parameters, since the planets
migrate to the edge of the cavity.  Additionally, they not only simulated full
grown planets but also migrating planetary cores. After the cores reached their
final positions, gas accretion and dispersion of the gas disc were initiated.
However the final orbital parameters were comparable, independent of the
migration scenario. For all studied systems, they found a set of disc parameters
which produced the closest approximation to the observed values. But since they
could not precisely reproduce the observed orbital parameters, they suggested
more sophisticated disc models.

\citet{2014A&A...564A..72K} simulated planets around Kepler-38 in isothermal as
well as radiative discs. Their radiative disc models include viscous heating,
vertical cooling, and radiative diffusion in the midplane of the disc. In both
models, the planet migrated as expected to the edge of the inner cavity. In
their isothermal models, the inner gap was smaller than the observed orbit of
the planet and therefore the planet migrated too close to the binary. However,
this small inner cavity was a result of their overly large inner radius of the
computational domain with $R_\mathrm{min} = 1.67\,a_\mathrm{bin}$. As shown in
\citet{2017A&A...604A.102T}, this inner radius should be of the order of the
binary separation. In the radiative case, the viscous heating produced a disc
with an even smaller inner cavity. Only by reducing the disc's mass and its
viscosity can a wider inner cavity and a final orbit of the planet close to the
observed location be achieved.

In a second paper, \citet{2015A&A...581A..20K} simulated the evolution of
planets in isothermal and radiative discs around the Kepler-34 system. Because
of the high eccentricity of the Kepler-34 binary, a disc with a large eccentric
inner cavity is created. Therefore the planets stop at a position far beyond the
observed location. Between the isothermal and radiative models, they could not
find large differences. Additionally, they observed an alignment of the planets'
orbit with the precessing inner gap.  

\citet{2017MNRAS.469.4504M} studied planets in self-gravitating circumbinary
discs around the systems Kepler-16, -34, and -35. Their main result is that for
very massive discs (five to ten times the minimum mass solar nebula, MMSN), the
disc's self gravity can shrink the inner cavity which allows the planet to
migrate further inward. This way they could achieve a better agreement between
their simulations of the Kepler-16 and Kepler-34 systems and the observed
values. In the case of Kepler-35b, the low mass of the planet prevented the
planet from migrating close to the observed location.

In this paper, we revisit the evolution of embedded planets in circumbinary
discs based on our new, refined disc models presented in
\citet{2017A&A...604A.102T}.  To study the migration of fully formed planets
through a circumbinary disc we carried out two dimensional (2D), isothermal,
viscous hydrodynamical simulations. The binary parameters were chosen according
to five selected Kepler systems.  Table~\ref{tab:systems} gives an overview of
binary and planet parameters of these systems. First, we simulated the
circumbinary disc without the presence of a planet to obtain the dynamical
properties of the inner cavity that we compare to our first study
\citep{2017A&A...604A.102T}. In order to detect purely numerical features and
strengthen our results, we used two grid codes, \textsc{Pluto} and
\textsc{Fargo3D}, that use different numerical methods, and compared their
results. 
\begin{table*}[ht]
    \caption{Circumbinary systems.}
    \label{tab:systems}
    \centering
    \begin{tabular}{c c c c c c c c c c}
        \hline\hline
        \noalign{\smallskip}
        System & 
        $M_\mathrm{A}\;[M_\sun]$ &
        $M_\mathrm{B}\;[M_\sun]$ & 
        $q_\mathrm{bin}$ &
        $a_\mathrm{bin}\;[\mathrm{au}]$ & 
        $e_\mathrm{bin}$ &
        $T_\mathrm{bin}\;[\mathrm{d}]$ &
        $m_\mathrm{p}\;[M_\mathrm{jup}]$ & 
        $\ap\;[\mathrm{au}]$ &
        $\ep$ \\
        \noalign{\smallskip}
        \hline
        \noalign{\smallskip}
        Kepler-16 & 0.69 & 0.20 & 0.29 & 0.22 & 0.16 & 41.079 & 0.333 & 0.7048 & 0.00685 \\
        \\
        Kepler-34 & 1.05 & 1.02 & 0.97 & 0.23 & 0.52 & 27.796 & 0.220 & 1.0896 & 0.182 \\
        \\
        Kepler-35 & 0.89 & 0.81 & 0.91 & 0.18 & 0.14 & 20.734 & 0.127 & 0.6035 & 0.042 \\
        \\
        Kepler-38 & 0.95 & 0.25 & 0.26 & 0.15 & 0.10 & 18.795 & < 0.384 & 0.4644 & < 0.032 \\
        \\
        Kepler-413 & 0.82 & 0.54 & 0.66 & 0.10 & 0.04 & 10.116 &0.211 & 0.3553 & 0.1181 \\
        \noalign{\smallskip}
        \hline
    \end{tabular}
    \tablefoot{The mass ratio is defined as $q_\mathrm{bin} =
    M_\mathrm{B}/M_\mathrm{A}$.}
    \tablebib{
    Kepler-16 \citep{2011Sci...333.1602D}, 
    Kepler-34 and -35 \citep{2012Natur.481..475W}, 
    Kepler-38 \citep{2012ApJ...758...87O},
    Kepler-413 \citep{2014ApJ...784...14K}.}
\end{table*}

For simulations with an embedded planet, we studied the evolution of the planet's
orbital elements as well as the planet's impact on the disc. In these
simulations, we varied the planet's mass and the mass of the disc, in order to
investigate the influence of the planet-to-disc mass ratio on the final orbital
parameters of the planet.

The paper is organised in the following way. In Sect.~\ref{sec:setup}, we
describe the physical and numerical setup of our simulations. In
Sect.~\ref{sec:disc_structure}, we discuss the disc structure prior to inserting
the planet and compare it to our earlier study. In Sect.~\ref{sec:planets}, we
investigate how planets of different mass migrate through these circumbinary
discs. We first discuss the general behaviour of migrating planets of different
mass using the example of Kepler-38, and then comment on the other four systems.
Our results are discussed and summarised in Sects.~\ref{sec:discussion} and
\ref{sec:summary}.

\section{Model setup}\label{sec:setup}
\subsection{Physical model}
To model the migration of planets in circumbinary discs, we use 2D,
isothermal, viscous hydrodynamical simulations. For the disc model, we follow the
setup described in~\citet[Sec. 2]{2017A&A...604A.102T}. This means we solve the
isothermal, vertically averaged Navier-Stokes equations with an external
potential $\Phi$ on a polar grid $(R, \varphi)$\footnote{We use the following
notation: $\vec{R}$ is the 2D position vector in the $x-y$-plane,
$\vec{R} = R \vec{\hat{e}_R}$}, centred at the barycentre of the binary.

We assume a locally isothermal temperature profile of $T \propto R^{-1}$ which
corresponds to a disc with constant aspect ratio $h = H/R$, where $H$ is the
height of the disc. In all simulations, we use $h = 0.05$.
Turbulent viscosity in the disc is modelled through an
$\alpha$-parameter\citep{1973A&A....24..337S}, with $\alpha = 0.01$.
We have chosen these parameters to be consistent with our first paper
but have run some exploratory simulations with smaller $h$ and different $\alpha$,
we comment on this in Sect.~\ref{sec:disc_structure} and the Discussion below.

The external potential $\Phi$ has the following form:
\begin{equation}\label{eq:potential}
    \Phi(\vec{R}) = -\sum_k \frac{G M_k}{ \sqrt{(\vec{R}-\vec{R}_k)^2
                                           +(\varepsilon H)^2}}
                    +\vec{a}_\mathrm{com} \cdot \vec{R}
,\end{equation}
with $k$ being an index running over both binary components and the planet
($k\in \{\mathrm{A}, \mathrm{B}, \mathrm{p} \}$). The subscripts $\mathrm{A}$ 
and $\mathrm{B}$ stand for the primary and the
secondary star, and the subscript $\mathrm{p}$ for the planet.  The smoothing factor
$\varepsilon H$ is used to avoid singularities and to account for the correct
treatment of a vertically extended three-dimensional (3D) disc in our 2D
thin-disc approximation \citep{2012A&A...541A.123M}.  In all our simulations, we
use $\varepsilon = 0.6$. The disc height $H$ is evaluated at the location of the
cell $H = h R$. Since our coordinate system is located at the centre of mass of
the binary, which is accelerated by the planet and the disc, we are not in an
inertial reference system. This acceleration of the coordinate system,
$\vec{a}_\mathrm{com}$, gives rise to the indirect term $\vec{a}_\mathrm{com}
\cdot \vec{R}$ in equation~\eqref{eq:potential}, where $\vec{a}_\mathrm{com}$ is
given by
\begin{equation}\label{eq:acc_com}
    \begin{split}
        \vec{a}_\mathrm{com} &= \frac{ M_\mathrm{A} \vec{a}_\mathrm{A}
                                      +M_\mathrm{B} \vec{a}_\mathrm{B}}
                                     {M_\mathrm{bin}} \\
                             &= \frac{1}{M_\mathrm{bin}}
                                \sum_{k\in\{\mathrm{A},\mathrm{B}\}}
                                \frac{G M_\mathrm{p} M_k}{|\vec{R}_\mathrm{p}-\vec{R}_k|^3}
                                (\vec{R}_\mathrm{p}-\vec{R}_k) \\
                             &\quad +\frac{1}{M_\mathrm{bin}}
                                     \sum_{k\in\{\mathrm{A},\mathrm{B}\}}
                                     \int_\mathrm{disc}
                                     \frac{G M_k \Sigma(\vec{R}') \dd V'}
                                          {\left[ (\vec{R'}-\vec{R}_k)^2 
                                                 +(\varepsilon H)^2\right]^{3/2}}
                                     (\vec{R}'-\vec{R}_k)
    \end{split}
,\end{equation}
with the mass of the binary $M_\mathrm{bin} = M_\mathrm{A} + M_\mathrm{B}$.  The first term in
equation~\eqref{eq:acc_com} is the acceleration of the centre of mass due to the
planet and the second term is the acceleration due to the disc. 

The equation of motion for the binary components and the planet are given by
\begin{equation}\label{eq:eq_of_motion}
    \begin{split}
        \frac{\mathrm{d}^2 \vec{R}_k}{\mathrm{d}t^2} 
        = &- \sum_{\ell\neq k} 
             \frac{G M_\ell}{|\vec{R}_k - \vec{R}_\ell|^3}
             (\vec{R}_k - \vec{R}_\ell)\\
          &+ \int_\mathrm{disc} 
             \frac{G \Sigma(\vec{R}') \dd V'}
                  {\left[(\vec{R}'-\vec{R}_k)^2+(\varepsilon H)^2\right]^{3/2}}
             (\vec{R}' - \vec{R}_k)\\
          &- \vec{a}_\mathrm{com} \,.
    \end{split}
\end{equation}
Again the height of the disc is evaluated at the location of the cell $H = h R'$
\citep{2012A&A...541A.123M}. Equations~(\ref{eq:acc_com}) and
(\ref{eq:eq_of_motion}) show the most general case in which the binary is
accelerated by the planets and the disc. For disc-only simulations in Sect.
\ref{sec:disc_structure}, the back-reaction of the disc on the binary is not
considered and therefore the disc contributions in Eqs.~(\ref{eq:acc_com}) and
(\ref{eq:eq_of_motion}) are omitted.

\subsection{Initial conditions}
The initial disc density is given by 
\begin{equation}\label{eq:sigma0}
    \Sigma(t=0) = f_\mathrm{gap}\,
                  \Sigma_\mathrm{ref}\, 
                  \left(\frac{R}{a_\mathrm{bin}} \right)^{-1.5} \,,
\end{equation}
where $f_\mathrm{gap}$ models the expected cavity created by the binary
\citep{1994ApJ...421..651A,2002A&A...387..550G}
\begin{equation}
    f_\mathrm{gap} = \left[1+\exp{\left(-\frac{R-R_\mathrm{gap}}
                                              {\Delta R}\right)} \right]^{-1} \,,
\end{equation}
with $R_\mathrm{gap} = 2.5\,a_\mathrm{bin}$, and $\Delta R =
0.1\,R_\mathrm{gap}$.  For the migration process, the total mass of the disc is
an important quantity, since it directly influences the migration speed of the
planet. In all our models, we choose an initial disc mass of $M_\mathrm{disc} =
0.01\,M_\mathrm{bin}$. With this choice, the reference surface density in
eq.~\eqref{eq:sigma0} can then be calculated:
\begin{align}
    M_\mathrm{disc} &= \int_0^{2\pi} \int_{R_\mathrm{min}}^{R_\mathrm{max}}
                       \Sigma(t=0) R \dd R \dd \varphi \\
                    &= 2\pi\Sigma_\mathrm{ref} 
                       \int_{R_\mathrm{min}}^{R_\mathrm{max}}
                       \left[1+\exp\left\{-\frac{\frac{R}{a_\mathrm{bin}} - 2.5}
                        {0.25} \right\} \right]^{-1}
                        \left(\frac{R}{a_\mathrm{bin}} \right)^{-1.5} R \dd
                        R \,.
\end{align}
Introducing the dimensionless variable $u = R/a_\mathrm{bin}$ (with
$u_{\mathrm{min}/\mathrm{max}} = R_{\mathrm{min}/\mathrm{max}}/a_\mathrm{bin}$) 
gives
\begin{equation}
    M_\mathrm{disc} = 2\pi \Sigma_\mathrm{ref} a_\mathrm{bin}^2 
                      \underbrace{\int_{u_\mathrm{min}}^{u_\mathrm{max}} 
                      \left[1 + \exp\left\{-\frac{u-2.5}{0.25} \right\} 
                      \right]^{-1} u^{-0.5} \dd u}_{= I(u_\mathrm{min},
                      u_\mathrm{max})}\,.
\end{equation}
Finally the reference density is given in our case in code units\footnote{In all
simulations, we use the following code units: $R_0 = a_\mathrm{bin}$, $M_0 =
M_\mathrm{bin}$, $t_0 = \sqrt{R_0^3/(G M_0)}$ and $\Sigma_0 = M_0/R_0^2$, with
the gravitational constant $G$.} with $u_\mathrm{min} = 1$ and $u_\mathrm{max} =
40$ by
\begin{equation}
    \frac{\Sigma_\mathrm{ref}}{\Sigma_0} = \frac{0.01}{2\pi I(1, 40)} 
                                         = 1.67535\times 10^{-4}\,;
\end{equation} 
the integral $I(1,40)$ was evaluated numerically.

The initial radial velocity is set to zero $u_R(t=0) = 0$ and the initial
azimuthal velocity is set to the local Keplerian velocity
$u_\mathrm{\varphi}(t=0) = \sqrt{G M_\mathrm{bin}/ R}$. 

\subsection{Numerics}\label{ssec:numerics}
The grid spans from $R_\mathrm{min} = 1\,a_\mathrm{bin}$ to $R_\mathrm{max} =
40\,a_\mathrm{bin}$ in the radial direction with a logarithmic spacing and from
$0$ to $2\pi$ in the azimuthal direction with a uniform spacing. In all
simulations, we use a resolution of $684 \times 584$ grid cells. At the inner
edge, we use a zero-gradient boundary condition ($\partial/\partial R = 0$) for
density, radial velocity, and angular velocity $\Omega_\varphi = u_\varphi/R$. We
further allow only outflow and no inflow into the computational domain. At the
outer boundary, we set the azimuthal velocity to the local Keplerian velocity.
For the density and radial velocity, we use a damping boundary condition in the
range from $(1-f)R_\mathrm{max}$ to $R_\mathrm{max}$ (with $f=0.1$) as described
in~\citet{2006MNRAS.370..529D}. This means we damp a quantity $x$ (here surface
density $\Sigma$ or radial velocity $u_R$) to its initial value $x^0$ according
to
\begin{equation}
    x^{n+1} = x^n - (x^n - x^0) \frac{\Delta t}{\tau} \mathcal{P}(R)
,\end{equation}
where $\Delta t$ is the current time-step, $\tau$ is the damping time scale and
equal to one tenth of the orbital period at the outer boundary $\tau = 0.1
\times 2\pi \sqrt{R_\mathrm{max}^3/(G M_\mathrm{bin})}$. The damping function
$\mathcal{P}(R)$ is a quadratic polynomial which fulfils the following
conditions $\mathcal{P}([1-f] R_\mathrm{max}) = 0.0$,
$\mathcal{P}(R_\mathrm{max}) = 1.0$ and $\mathcal{P}'([1-f]R_\mathrm{max}) =
0.0$. Therefore,
\begin{equation}
    \mathcal{P}(R) = \frac{1}{f^2} \left(\frac{R}{R_\mathrm{max}} \right)^2
                    +\frac{2(f-1)}{f^2} \frac{R}{R_\mathrm{max}} 
                    +\frac{(1-f)^2}{f^2}\,.
\end{equation}
In the azimuthal direction, we use periodic boundary conditions.

For stability reasons, especially in the inner cavity where the density drops to
very low values, we use a global density floor,
\begin{equation}
    \Sigma_\mathrm{floor} = 10^{-6}\,\Sigma_\mathrm{ref}
,\end{equation}
in all our simulations. Without a density floor, some hydrodynamical codes,
especially \textsc{Pluto}, tend to produce negative densities and abort the
calculation. Test simulations with lower floors did not show any differences in
disc dynamics, but negative densities and code abortions occurred more
frequently. Simulations with density floors higher than
$10^{-6}\,\Sigma_\mathrm{ref}$ did not agree with the lower floor results, hence
our choice for $\Sigma_\mathrm{floor}$.

In this paper, we use two hydrodynamical codes, namely \textsc{Pluto}
\citep{2007ApJS..170..228M}\footnote{We use a modified version of \textsc{Pluto}
4.2 which runs on GPUs and is also capable of solving a N-body system.} and
\textsc{Fargo3D} \citep{2016ApJS..223...11B}. For a quick overview of the
numerical options used in our simulations see
\citet[Sec. 3.1]{2017A&A...604A.102T}. Some details about the N-body
solver implemented in \textsc{Pluto} are shown in Appendix~\ref{sec:pluto_nbody}.

\textsc{Fargo3D} uses an artificial von Neumann-Richtmyer viscosity to smooth
out shocks. The artificial viscosity follows the \textsc{Zeus}
implementation~\citep{1992ApJS...80..753S}. The number of cells over which the
shock is spread can be adjusted through a constant $C_\mathrm{av}$ (in the
\textsc{Zeus} paper, this constant is referred to as $C_2$). \textsc{Fargo3D} sets this
constant by default to $C_\mathrm{av} = 1.41$. As discussed in
Appendix~\ref{sec:comparison_with_fargo}, to obtain a good agreement with
\textsc{Pluto} a lower artificial viscosity is needed. Therefore, we have chosen
$C_\mathrm{av} = 0.5$ for all \textsc{Fargo3D} simulations shown in this paper.

\begin{figure}
    \centering
    \resizebox{\hsize}{!}{\includegraphics{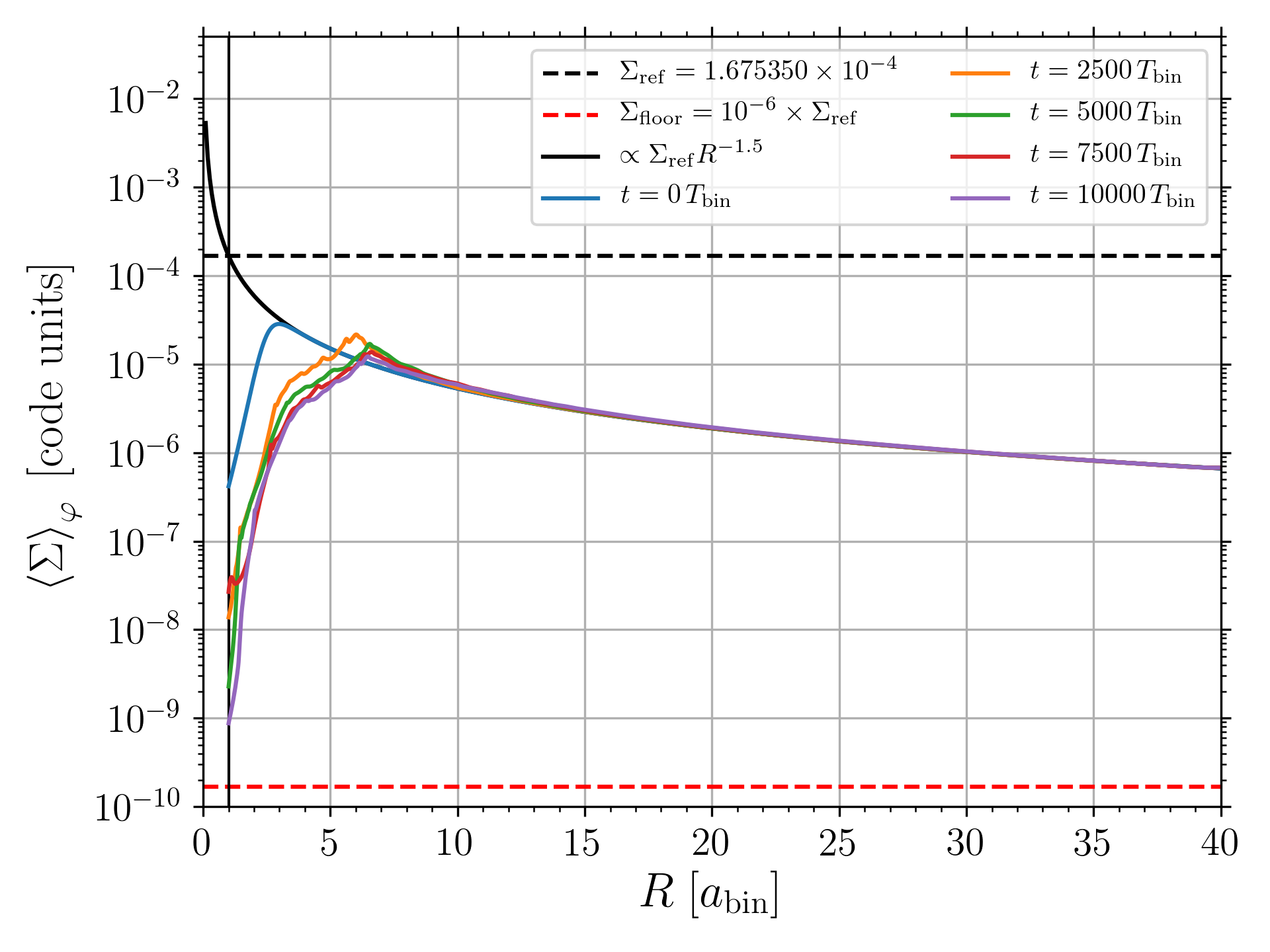}}
    \caption{Azimuthally averaged surface density at various times for the
    Kepler-38 system calculated with \textsc{Pluto}. The blue curve shows the
    initial density distribution. The black vertical line marks the
    $1\,a_\mathrm{bin}$ location where the density would reach the reference
    density if we had not imposed an initial gap (blue curve). The dashed
    horizontal lines mark the reference density (in black) and the density floor
    (in red).}
    \label{img:k38_floor_pluto}
\end{figure}
\begin{table}
    \caption{Fixed numerical and physical parameters of our simulations.}
    \label{tab:parameters}
    \centering
    \begin{tabular}{ll}
        \hline\hline
        \noalign{\smallskip}
        Physical parameters &  \\
        \noalign{\smallskip}
        \hline
        \noalign{\smallskip}
        $h$ & 0.05 \\
        $\alpha$ & 0.01 \\
        $M_\mathrm{disc}$ & $0.01\,M_\mathrm{bin}$\\
        \noalign{\smallskip}
        \hline
        \noalign{\medskip}
        Numerical parameters &  \\
        \noalign{\smallskip}
        \hline
        \noalign{\smallskip}
        $\mathrm{Resolution}$ & $684 \times 584$ \\
        $R_\mathrm{min}$ & $1\,a_\mathrm{bin}$\\
        $R_\mathrm{max}$ & $40\,a_\mathrm{bin}$\\
        $\Sigma_\mathrm{floor}$ & $10^{-6}\,\Sigma_\mathrm{ref}$ \\
        $\varepsilon$ & $0.6$ \\
        \noalign{\smallskip}
        \hline
    \end{tabular}
\end{table}
Figure~\ref{img:k38_floor_pluto} summarises our setup using the Kepler-38 system
as an example.  Displayed are various snapshots of the azimuthally averaged
surface density in order to show that the density floor does not directly
interfere with the disc evolution. The density snapshot are from a
\textsc{Pluto} simulation of Kepler-38 without a planet. The planets are then
embedded at $t=10000$ binary orbits into the disc.  In
Table~\ref{tab:parameters}, all numerical and physical parameters which were not
varied in our simulations are summarised.

\section{Circumbinary disc structure}\label{sec:disc_structure}
\begin{figure*}
    \centering
    \includegraphics[width=17cm]{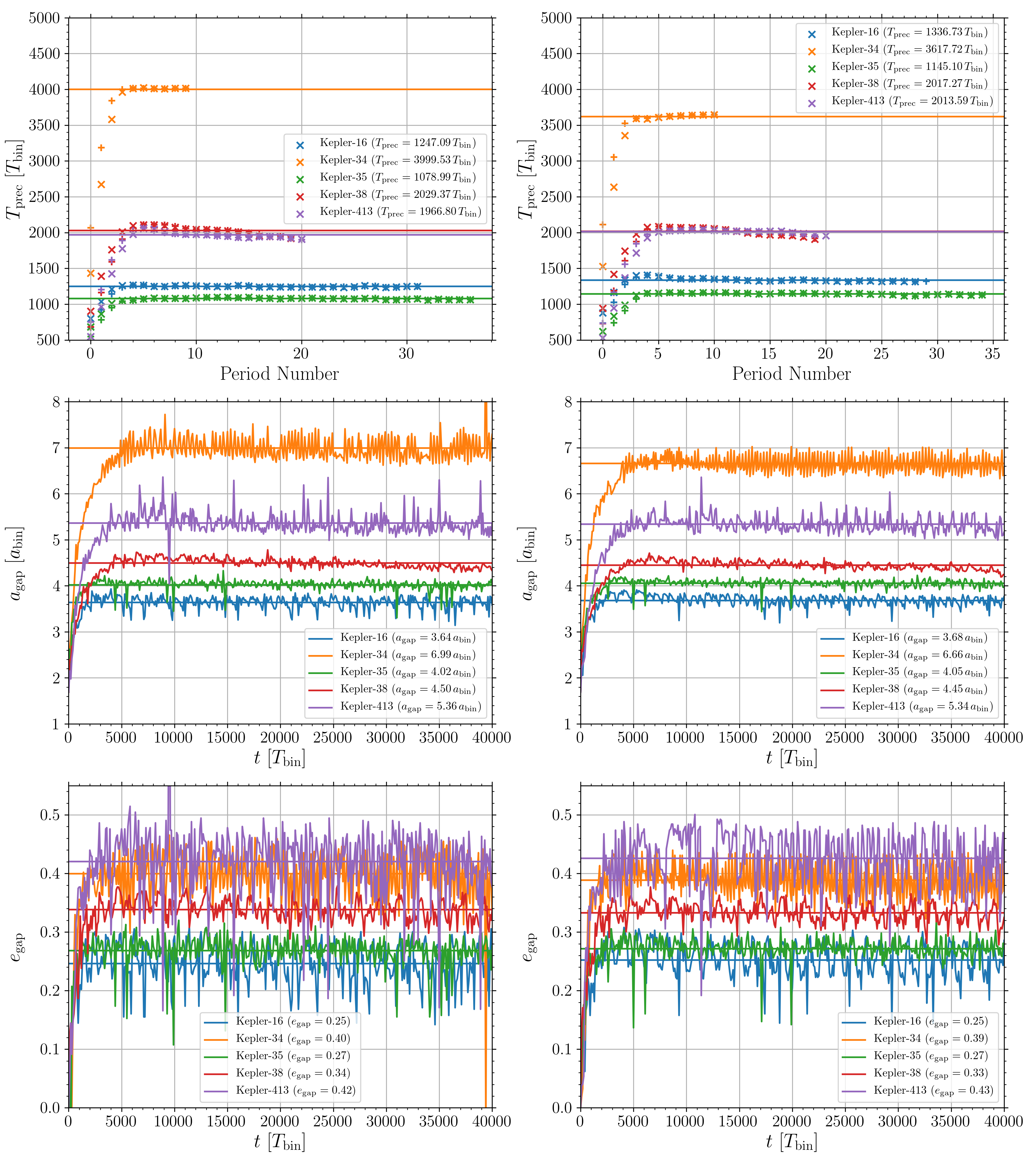}
    \caption{The precession period, semi-major axis, and eccentricity of the disc
    gap vs. time, for our five different systems. Disc models on the left are
    calculated with \textsc{Pluto} and disc models on the right are calculated
    with \textsc{Fargo3D}. The precession period, $T_\mathrm{prec}$, in the
    first row is given in terms of the corresponding period number.  The
    horizontal lines indicate the time average starting from \num{6000} binary
    orbits, roughly the time when the discs reached a quasi-steady state. See
    the text for an explanation of the meaning of the $+$ and $\times$ signs in
    the first row.}
    \label{img:disc_properties}
\end{figure*}
Before studying the evolution of embedded planets, we present our results on the
dynamical behaviour of the circumbinary disc around the five systems under 
investigation.  In Fig.~\ref{img:disc_properties}, we display the time evolution
over \num{40000} orbits of the binary of the disc's precession period as well as
the semi-major axis and the eccentricity of the disc's gap.  In the left column,
results from \textsc{Pluto} simulations are shown and in the right column
\textsc{Fargo3D} results are presented.

The semi-major axis and eccentricity of the gap are calculated with the help of
a fitted ellipse. The detailed fitting procedure is described in \citet[Sect.
5.1]{2017A&A...604A.102T}. The displayed data show some noise since we
calculated the semi-major axis and eccentricity of the disc only every 100
binary orbits.  The final estimates are time averages, starting at \num{6000}
binary orbits (roughly the time when all discs reached a quasi steady state).
The precession period is calculated in two ways: We searched for positive ($\pi
- \delta \rightarrow \pi + \delta, \delta \ll 1$) and negative ($\pi + \delta
\rightarrow \pi - \delta$) transitions in the $\varpi_\mathrm{disc}$-$t$-diagram
(for an example of a $\varpi_\mathrm{disc}$-$t$-diagram see
Fig.~\ref{img:disc_planet_peri} below).  The time between two subsequent
positive ($\times$ in Fig.~\ref{img:disc_properties}) or negative ($+$ in
Fig.~\ref{img:disc_properties}) transitions gives  the precession period of the
gap, which is shown in the first row of Fig.~\ref{img:disc_properties} ordered
by their number of occurrence.  To obtain a better estimate, we averaged over all
periods after the discs reached a quasi-steady state, which happens after about
\num{6000} binary orbits. The overall agreement between the results obtained
with \textsc{Pluto} and \textsc{Fargo3D} is very good; we discuss some
notable differences below.

\begin{figure}
    \centering
    \resizebox{\hsize}{!}{\includegraphics{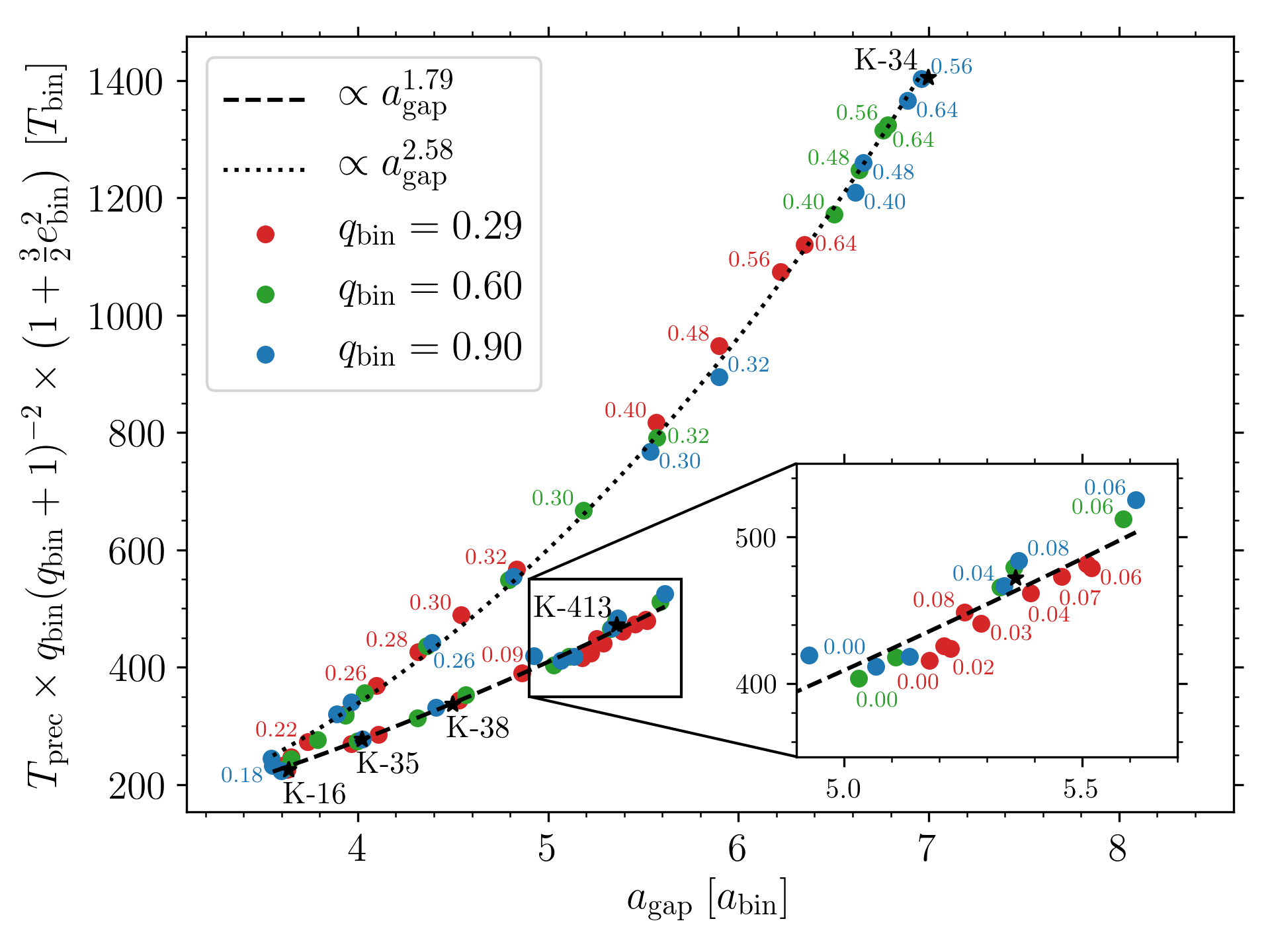}}
    \caption{Precession period of the inner gap, scaled by the factor
    $q_\mathrm{bin} / (q_\mathrm{bin} + 1)^2 \times \left(1 + \frac{3}{2}
    e_\mathrm{bin}^2 \right)$ to account for the different mass ratios and
    eccentricities of the binary stars, plotted against the semi-major axis of
    the gap. Different bullets correspond to numerical results for different
    binary eccentricities whereas different colours stand for different binary
    mass ratios. The black dashed and dotted lines are fits to all data points
    on the lower (dashed) and upper (dotted) branch. The five
    different Kepler systems considered in this paper are marked by the black
    stars.}
    \label{img:Tprec_vs_agap}
\end{figure}
Considering the different binary eccentricities in our systems, as noted in
Table~\ref{tab:systems}, we can place them in a $T_\mathrm{prec}$ versus
$a_\mathrm{gap}$ diagram; see Fig.~\ref{img:Tprec_vs_agap}.  In a previous study
we discovered a bifurcation with two separate branches depending on the binary
eccentricity \citep[Sect.~5.2]{2017A&A...604A.102T}.  In that study we found that
on the lower branch, starting with circular binaries, the precession period and
gap size decreases for increasing binary eccentricities. The minimum gap size
and precession period is reached at a critical binary eccentricity of
$e_\mathrm{crit} \approx 0.18$. From there, the upper branch starts on which both
quantities increase for increasing binary eccentricities. The vertical axis in
Fig.~\ref{img:Tprec_vs_agap} has been rescaled using the factor
\begin{equation}
    \frac{q_\mathrm{bin}}{(q_\mathrm{bin} + 1)^2}
    \times\left(1 + \frac{3}{2} e_\mathrm{bin}^2 \right) \,,
\end{equation}
to account for the different binary mass ratios and eccentricities. As shown in
\citet[Appendix B]{2017A&A...604A.102T} the precession period of the disc gap is
comparable to the precession period of a single particle around a binary,
\begin{equation}\label{eq:particle_tprec}
    T_\mathrm{prec} = \frac{4}{3} 
                      \frac{(q_\mathrm{bin} + 1)^2}{q_\mathrm{bin}}
                      \left(\frac{a_\mathrm{p}}{a_\mathrm{bin}} \right)^{7/2}
                      \frac{\left(1-e_\mathrm{p}^2\right)^2}
                           {\left(1 + \frac{3}{2} e_\mathrm{bin}^2\right)}
                      T_\mathrm{bin}\,.
\end{equation}
Our scaling factor does not include the gap eccentricity term
$\left(1-e_\mathrm{gap}^2\right)^2$, 
in order to allow for a clear relation between binary and individual disc properties, here 
$T_\mathrm{prec}$ versus $a_\mathrm{gap}$.
From Fig. 4, it is, in principle, possible to construct a relation
between $e_\mathrm{gap}$ and $a_\mathrm{gap}$ and add this to the scaling in Fig. 3. 
Using $e_\mathrm{gap}$ in the scaling for Fig.~3 does not change it qualitatively, 
only the exponents are different then.

Using this scaling, the numerically obtained disc parameters can be placed into
the diagram for all systems, which lie indeed very close to the discovered
bifurcation curve of \citet{2017A&A...604A.102T}.  While Kepler-16 and Kepler-35
are located close to the branching point, the systems Kepler-38 and Kepler-413
lie on the lower branch.  The Kepler-34 system, which has a very large gap with a
slow precession, is located at the end of the upper branch in Fig.~\ref{img:Tprec_vs_agap}.

Fits to the upper and lower branch data points show that the scaled
precession period is proportional to $\propto a_\mathrm{gap}^{1.79}$ on the
lower branch and $\propto a_\mathrm{gap}^{2.58}$ on the upper branch. These
exponents are lower than the exponent $7/2$ expected from the test particle
precession period relation, eq.~\eqref{eq:particle_tprec}. This deviation represents
the fact that gap's dynamical behaviour is determined by hydrodynamical effects and
does not fully behave like a free particle around a binary.

To generate Fig.~\ref{img:Tprec_vs_agap}, we did not use the data from
\citet{2017A&A...604A.102T} but generated all points for the different binary
mass ratios and eccentricities using our new improved numerical setup as described above, in
particular a smaller $R_\mathrm{min}$. The red points in the figure are all
calculated for the Kepler-16 binary mass ratio of $q_\mathrm{bin}=0.29$, and the
green and blue bullets correspond to mass ratios of $q_\mathrm{bin}=0.60$ and
$q_\mathrm{bin}=0.90$. We also set up simulations series for these high mass
ratios since Kepler-34 also has a high mass ratio of $q_\mathrm{bin} = 0.97$. 
For binary
eccentricities,  $e_\mathrm{bin} \geq 0.06,$ we observe the same behaviour as in
our earlier study: upon increasing $e_\mathrm{bin}$ , the points move along the
lower branch towards smaller $a_\mathrm{gap}$ and $T_\mathrm{prec}$ until the
critical value $e_\mathrm{crit} \approx 0.18$ is reached, from which both
$T_\mathrm{prec}$ and $a_\mathrm{gap}$ increase with increasing
$e_\mathrm{bin}$. However,
for binary eccentricities smaller than $e_\mathrm{bin} = 0.06,$ we observe a new
phenomenon. Starting from circular binaries, the gap size as well as the
precession period first increase until a maximum is reached for $e_\mathrm{bin}
= 0.06$; see inlay in Fig.~\ref{img:Tprec_vs_agap}.  Increasing the
binary eccentricity past $e_\mathrm{bin} = 0.06$ leads to smaller gaps and lower
precession periods again, following the known trend. From $e_\mathrm{bin} =
0.0$ till $e_\mathrm{bin} = 0.08,$ a loop-like structure is created. 
In \citet{2017A&A...604A.102T}, we did not notice this loop-type behaviour for
small $e_\mathrm{bin}$ because we only simulated $e_\mathrm{bin} = 0.0$ and
$e_\mathrm{bin} = 0.08$ and no values in between.  Using a smaller
$R_\mathrm{min}  = a_\mathrm{bin}$ and a better coverage of small
$e_\mathrm{bin}$ values allowed us to discover this new, complex behaviour.
Although this constitutes a very exciting new finding, we defer further
investigation to a subsequent study. We only note here that, as shown in the
inlay of Fig.~\ref{img:Tprec_vs_agap}, this loop-like structure is also present
for higher mass ratios of $q_\mathrm{bin}=0.60$ and $q_\mathrm{bin}=0.90$ confirming
the generality of our results.

\begin{figure}
    \centering
    \resizebox{\hsize}{!}{\includegraphics{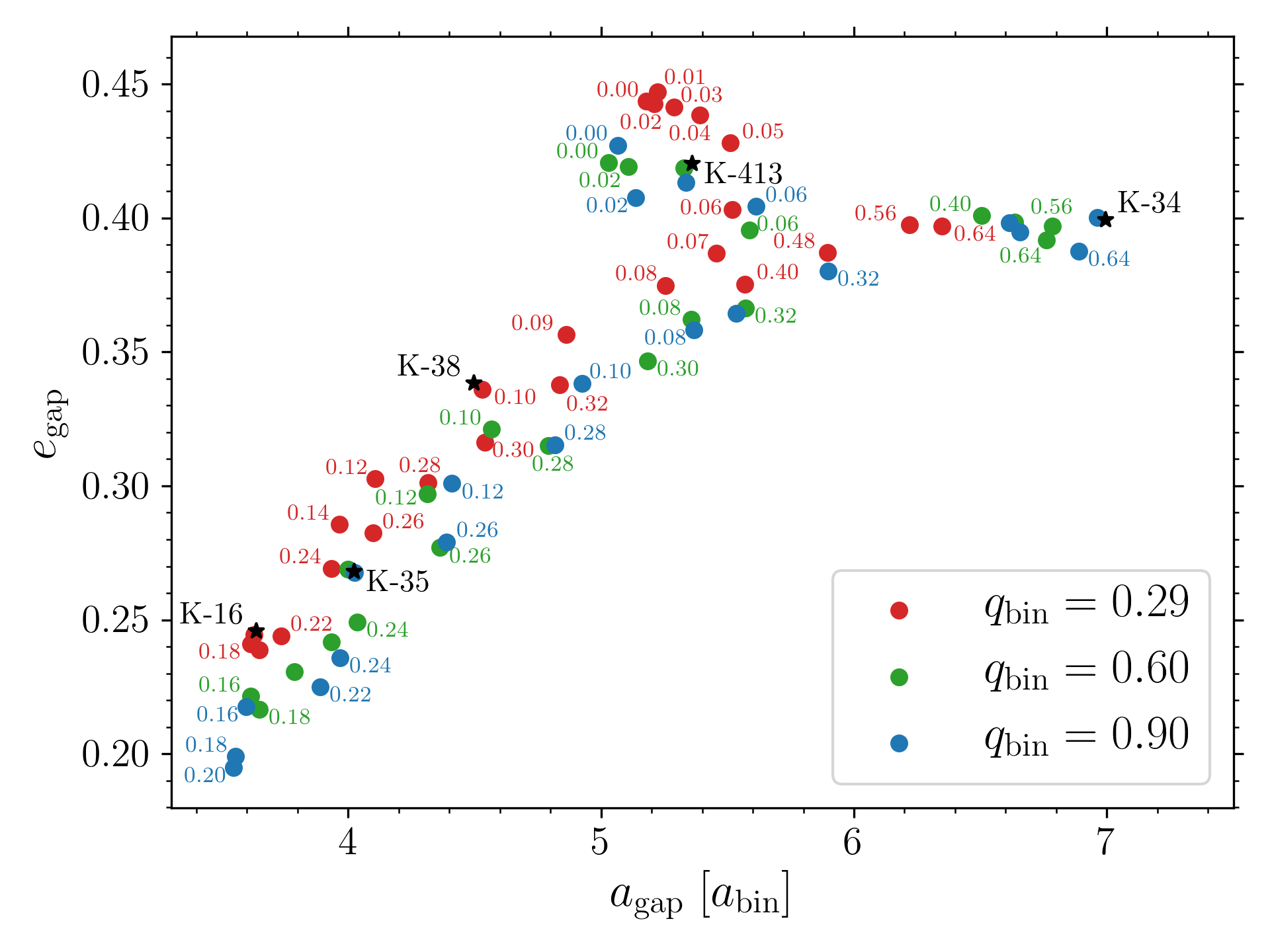}}
    \caption{Gap eccentricity plotted against the semi-major axis of the
    gap. Different bullets correspond to numerical results for different binary
    eccentricities whereas different colours stand for different binary mass
    ratios. The five different Kepler systems considered in this paper are
    marked by the black stars.}
    \label{img:egap_vs_agap}
\end{figure}
Figure~\ref{img:egap_vs_agap} shows the correlation of gap eccentricity and gap
size for different binary eccentricities as well as different binary mass
ratios. Discs around binaries with eccentricities close to zero create the most
eccentric gaps, while binaries with eccentricities close to the branching point
have the smallest and least eccentric gaps. High eccentric binaries create the
largest gaps which are also highly eccentric.

Figures \ref{img:Tprec_vs_agap} and \ref{img:egap_vs_agap} have been 
generated from models using our standard aspect ratio and viscosity. 
Exploratory simulations with different $h$ (0.03, 0.04) and $\alpha$ (0.001, 0.05) 
show the same trend as described in \citet{2017A&A...604A.102T}.
Increasing the viscosity leads to smaller, more circular
gaps with shorter precession periods, and the same trend holds for smaller aspect ratios.
Due to the neglect of the disc backreaction on the binary for these
simulations without embedded planet, the results displayed in Figs.
\ref{img:Tprec_vs_agap} and \ref{img:egap_vs_agap} refer to the zero disc mass
limit.  In our case of small disc mass ($M_\mathrm{disc} =
0.01\,M_\mathrm{bin}$) this is a very good approximation. For a test simulation
for our standard model (Kepler-38), which included disc backreaction, the change
in $e_\mathrm{bin}$ over the first 50000 binary orbits was less than 2\%.

Table~\ref{tab:disc_properties} summarises the gap properties of the five
systems as calculated using the \textsc{Pluto} code.  Although the systems
studied in this paper have mass ratios different from $q_\mathrm{bin}=0.29$,
which is the value of Kepler-16 investigated in \citet{2017A&A...604A.102T}, the
general behaviour is in good agreement with our earlier study. The most circular
system (Kepler-413) produces the most eccentric gap, with an eccentricity of
$e_\mathrm{gap} = 0.43$, while the largest gap is formed by the most eccentric
binary (Kepler-34), which also has the longest precession period of
$T_\mathrm{prec}=4000\,T_\mathrm{bin}$.  Systems close to the branching point
(Kepler-16 and Kepler-35) have the smallest gaps with the shortest precession
periods.  Kepler-35 has the shortest precession period of
$T_\mathrm{prec}=1079\,T_\mathrm{bin}$, since the mass ratio of Kepler-35 is,
with $q_\mathrm{bin} = 0.91,$ almost equal to one, and higher mass ratios lead to
shorter precession periods \citep{2017A&A...604A.102T}. Since the mass ratio
does not change the gap size significantly, Kepler-16, which has an eccentricity
closer to the critical eccentricity than Kepler-35, has the smallest gap of all
systems considered in this paper. As already seen in our earlier study we find
that the size of the gap is directly correlated with the precession period of
the gap, for all systems under investigation.  This behaviour is expected from
single particle trajectories and supported by the used scaling for
$T_\mathrm{prec}$ in Fig.~\ref{img:Tprec_vs_agap}.  Comparing with
Table~\ref{tab:systems}, one can see that the two systems with the highest gap
eccentricity (Kepler-34 and -413) also possess the planets with the highest
eccentricity. 
\begin{table}
    \caption{Disc properties of the five systems obtained with \textsc{Pluto}
    simulations.}
    \label{tab:disc_properties}
    \centering
    \begin{tabular}{l c c c}
        \hline\hline
        \noalign{\smallskip}
        System & 
        $T_\mathrm{gap}\;[T_\mathrm{bin}]$ &
        $a_\mathrm{gap}\;[a_\mathrm{bin}]$ & 
        $e_\mathrm{gap}$ \\
        \noalign{\smallskip}
        \hline
        \noalign{\smallskip}
        Kepler-16 & 1247.09 & 3.64 & 0.25 \\ 
        \\
        Kepler-34 & 3999.53 & 6.99 & 0.40 \\
        \\
        Kepler-35 & 1078.99 & 4.02 & 0.27 \\
        \\
        Kepler-38 & 2029.37 & 4.50 & 0.34 \\
        \\
        Kepler-413 & 1966.80 & 5.36 & 0.42 \\
        \noalign{\smallskip}
        \hline
    \end{tabular}
\end{table}

Concerning the outcome of the different numerical methods, the best agreement
between \textsc{Pluto} and \textsc{Fargo3D} is obtained for Kepler-38 and -413;
both systems are on the lower branch and far away from the branching point. For
our two systems close to the branching point (Kepler-16 and -35),
\textsc{Fargo3D} produces slightly higher precession periods than
\textsc{Pluto}, however the gap size and eccentricity match very well.  The
largest deviations between the two codes can be seen for the Kepler-34 system,
which produces the largest gap. Here the precession period deviates by roughly
ten percent and the gap size by five percent between the two codes.

Overall, the static disc parameters (gap size and eccentricity) are in good
agreement between the two codes for all systems. However, for the dynamical
parameter (precession period of the gap) the codes can deviate slightly,
depending on the observed system. This could be a consequence of the very low
precession rate, especially for Kepler-34, where the gap is nearly stationary
with respect to the inertial frame.  
\begin{figure}
    \centering
    \resizebox{\hsize}{!}{\includegraphics{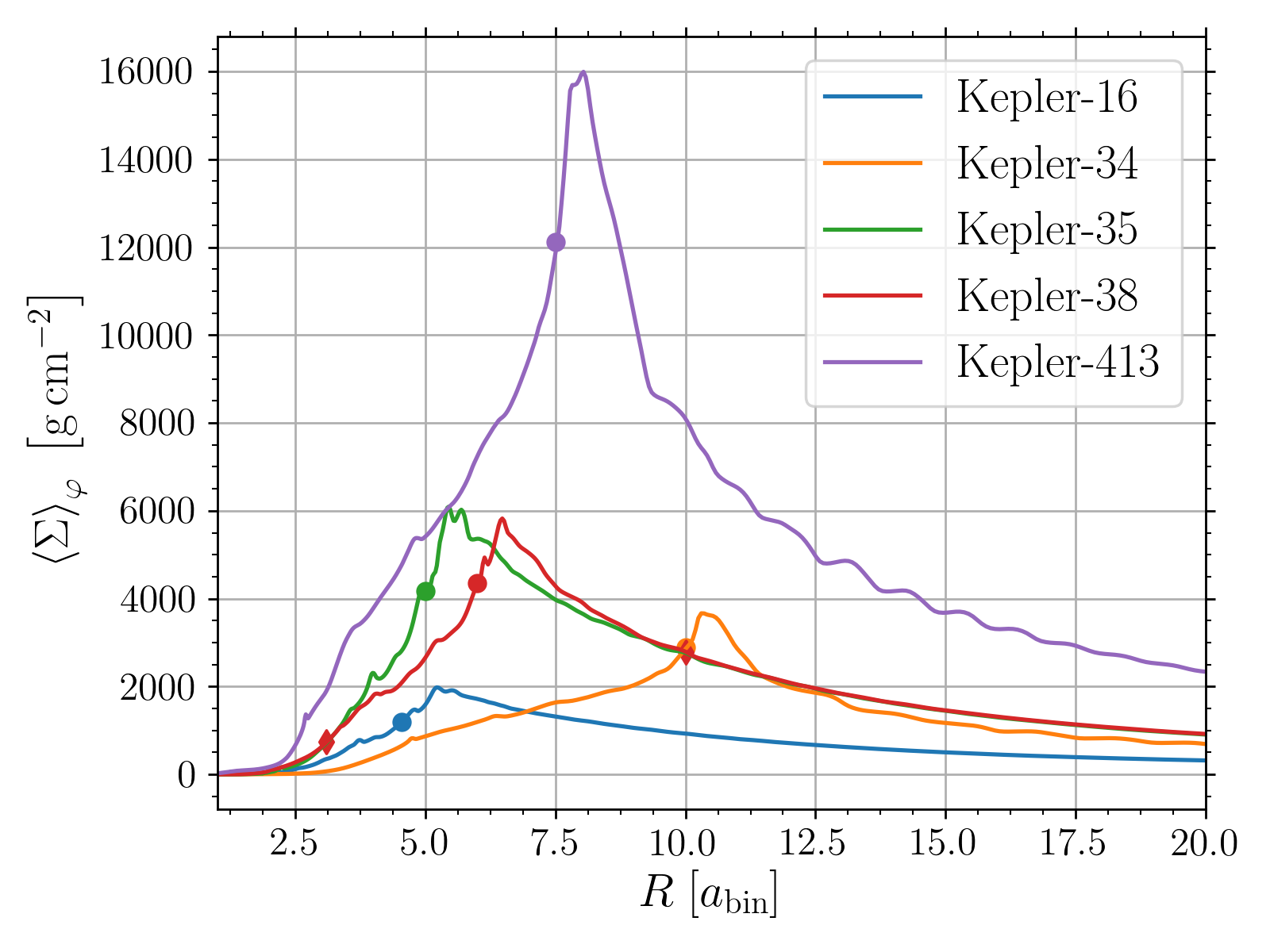}}
    \caption{Azimuthally averaged density profiles after \num{10000} binary
    orbits for the five different Kepler systems. The dots mark the initial
    planet position.}
    \label{img:sigma_prof_init_pos}
\end{figure}
\section{Planets in circumbinary discs}\label{sec:planets}
In this section, we describe the evolution of planets in the circumbinary
discs around our five systems under investigation. In the first part, we give an
account of the general behaviour using the sample system Kepler-38, and then
describe the specifics of the other systems below. Before inserting the planets
into the discs, we evolved the circumbinary discs for all systems without a
planet for $\num{10000}$ binary orbits such that a quasi steady state was
reached; see Fig.~\ref{img:disc_properties}. We then inserted the planet into the
disc and kept it on a fixed circular orbit for another \num{2000} binary orbits,
so that the disc could adjust to the presence of the planet. After that we
switched on the back-reaction of the disc onto the planet, meaning the planet
was allowed to migrate freely through the disc. We also switch on the
back-reaction of the disc onto the binary which leads to a very slow increase of
the binary eccentricity and a slow decrease of the binary separation,
accompanied with a small reduction of the binary orbital period,
$T_\mathrm{bin}$.  When we refer to $T_\mathrm{bin}$, we always mean the
initial, unperturbed binary period $T_\mathrm{bin} = 2\pi
\sqrt{a^3_\mathrm{bin}/(G M_\mathrm{bin})}$, calculated using the observed
values summarised in Table~\ref{tab:systems}. The back-reaction from the planet
and the disc causes also a precession of the binary, which is much slower
compared to the precession of the inner gap and the planet's orbit.  The
azimuthally averaged density profiles of the disc for all five systems after
$\num{10000}\,T_\mathrm{bin}$ is displayed in Fig.~\ref{img:sigma_prof_init_pos}
together with the initial positions of the embedded planets.

When simulating planets with the observed mass for our five systems, we
found two regimes with different dynamical behaviour of the planet.
To trigger these two regimes in all systems, we have
chosen planet-to-binary mass ratios $q$ from $q \approx 10^{-4}$, approximately the
observed planet-to-binary mass ratio in the Kepler-34 system (light case), up to
$q \approx 3.6 \times 10^{-4}$, approximately the observed planet-to-binary mass
ratio in the Kepler-16 system (massive case). This range of mass ratios leads
in the case of Kepler-34, -35, and -413 to planets which are way more massive
than the observed planet. Nonetheless, we simulated these high-mass planets to 
explore if the two regimes (here the massive case) can also be triggered in those systems as well.

In this section, all presented results were obtained by \textsc{Pluto}
simulations. In Appendix~\ref{sec:comparison_with_fargo}, we compare \textsc{Pluto} and
\textsc{Fargo3D} simulations of a circumbinary disc with an embedded planet.
%
%
\subsection{Kepler-38}\label{ssec:kepler38}
In this section, we use the Kepler-38 system as our reference system to
illustrate the general behaviour of a planet embedded in a circumbinary disc.
We have chosen Kepler-38 because for this system the best agreement between
\textsc{Pluto} and \textsc{Fargo3D} is obtained. Furthermore, we can compare our
results directly with \citet{2014A&A...564A..72K} who also looked at Kepler-38
applying a different numerical setup.  They used an initial density slope
proportional to $R^{-0.5}$ whereas we use $\Sigma \propto R^{-1.5}$ in this
study. However, test calculations have shown that the initial density profile has
no influence on the final quasi-steady state.  A more crucial difference is the
position of the inner boundary.  \citet{2014A&A...564A..72K} used an inner
radius of $R_\mathrm{min} = 1.67\,a_\mathrm{bin}$. However, as we have shown in
\citet{2017A&A...604A.102T} this radius is too large because it does not allow
the full disc dynamics to develop and results in very small and circular gap
edges.  For this reason, we have chosen here $R_\mathrm{min} = a_\mathrm{bin}$
for all systems. 
\begin{figure}
    \centering
    \resizebox{\hsize}{!}{\includegraphics{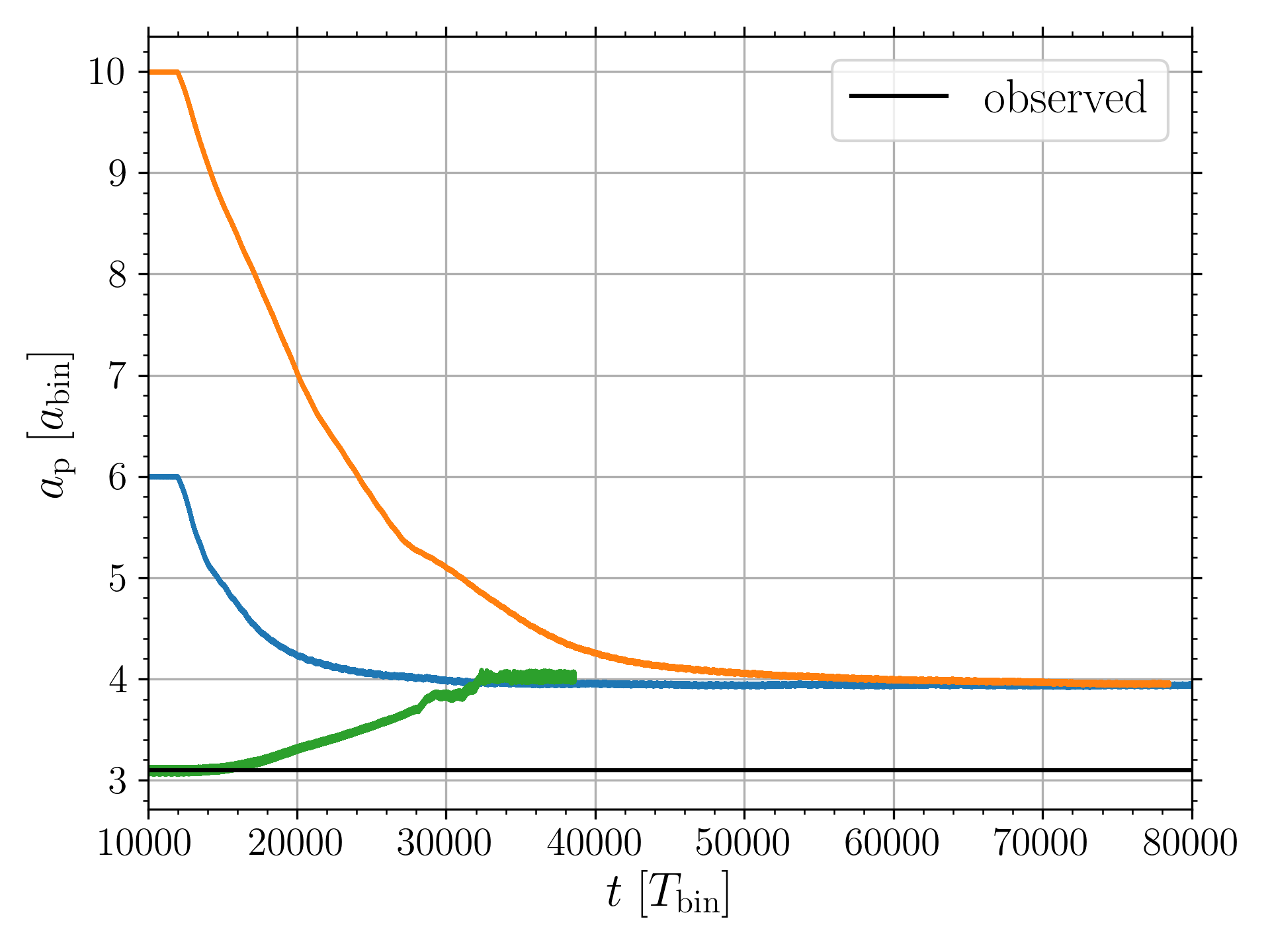}}
    \caption{Time evolution of the semi-major axis of a $m_\mathrm{p} =
    0.384\,M_\mathrm{jup}$ planet with different initial positions in a
    circumbinary disc around the Kepler-38 system. The black
    line shows the observed position of the planet.}
    \label{img:k38_init_planet_pos}
\end{figure}
%
\subsubsection{Initial planet position}\label{sssec:init_planet_pos}
In a first series of simulations, we investigate how the final stopping position
of the planet depends on its initial starting position. We simulated a
Saturn-like planet with a mass of $m_\mathrm{p} = 0.384\,M_\mathrm{jup}$, which is the
upper mass limit for Kepler-38b. Initially, we placed the planet at three
different locations inside the disc.  The red curve in
Fig.~\ref{img:sigma_prof_init_pos} shows the density profile for Kepler-38, and the
red symbols mark the three initial locations of the planet.  The outermost
starting point is at $R_\mathrm{p}(t=0) = 10\,a_\mathrm{bin}$, beyond the density
maximum, which is produced by the binary-disc interaction.  The second starting point was placed just
inside of the density maximum at $6.0\,a_\mathrm{bin}$, and finally the last starting point was placed at the
observed location $3.096\,a_\mathrm{bin}$.

Figure~\ref{img:k38_init_planet_pos} shows the time evolution of the semi-major
axis of the planet for the three initial starting points.  Independently of its
initial position, the planet migrates through the disc until it stops at a
distance of approximately $3.98\,a_\mathrm{bin} = 0.597\,\mathrm{au}$. The
observed semi-major axis of Kepler-38b is $a_\mathrm{obs} =
0.4644\,\mathrm{au}$. For the two outer starting points, the planet migrates
inward until it reaches the edge of the inner cavity, which is created by the
gravitational interaction of the binary on the disc. This behaviour is expected
since at this point the negative Lindblad torques, responsible for inward
migration, are balanced by the positive corotation torques and the planet
migration comes to rest \citep{2006ApJ...642..478M}. If we initially place the
planet at the observed location, which lies inside of the cavity, then the planet
starts to migrate outward to the edge of the disc where the torques are again in
balance (green curve in Fig.~\ref{img:k38_init_planet_pos}).  Such an
outward migration of a planet when starting inside of the cavity has been found
for the Kepler-34 system by \citet{2015A&A...581A..20K}. It is caused by the
transfer of positive angular momentum to the planet when interacting with the
eccentric inner disc.

\begin{figure}
    \centering
    \resizebox{\hsize}{!}{\includegraphics{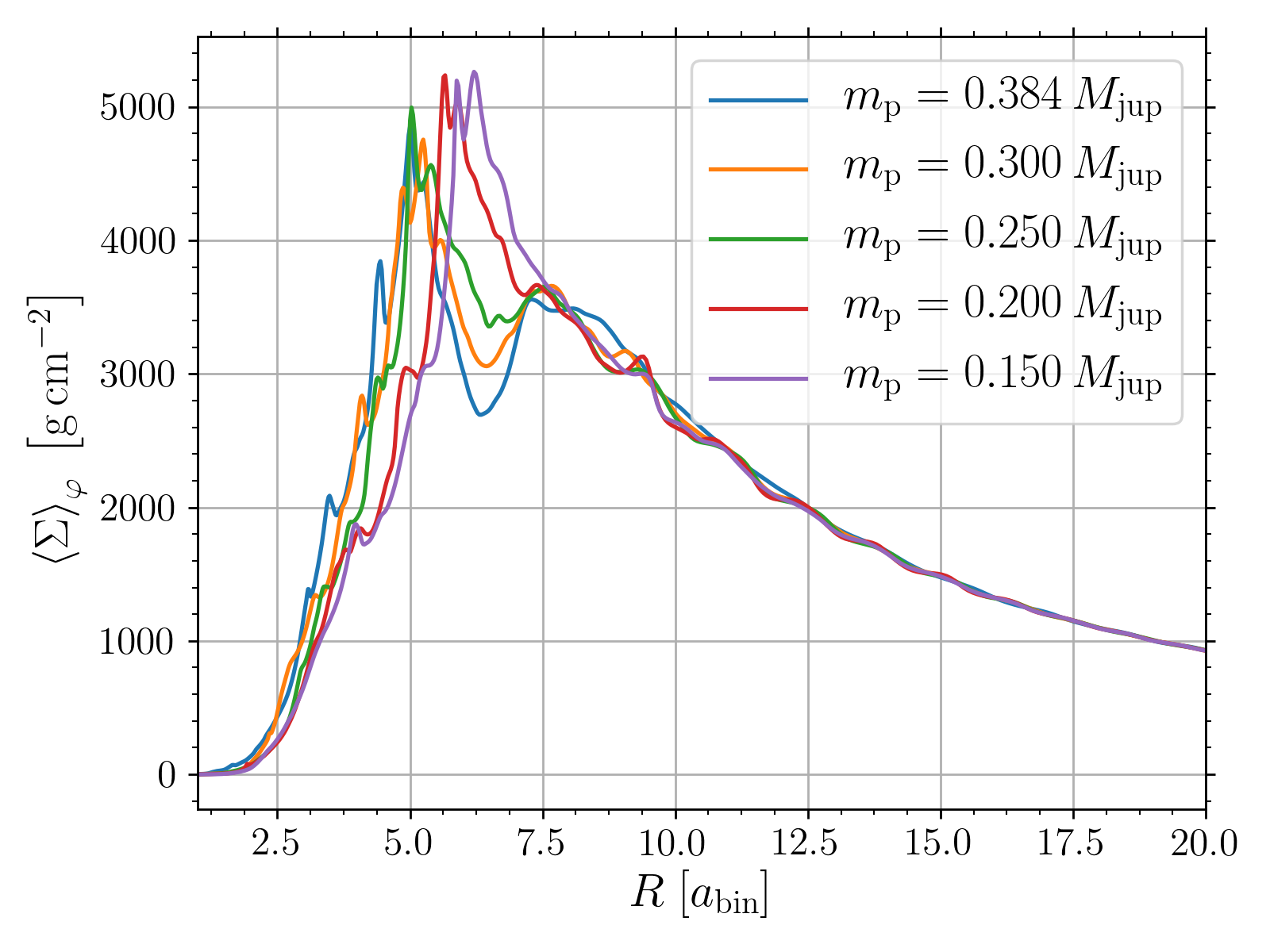}}
    \caption{Azimuthally averaged density profiles for the Kepler-38 disc
    with embedded planets after \num{12000} binary orbits. The planets were held
    on a fixed orbit ($a_\mathrm{p} = 6.0\,a_\mathrm{bin}$) for \num{2000} 
    binary orbits.}
    \label{img:k38_sigma_prof_gap}
\end{figure}

The independence of the final planet position from its initial conditions
suggest that the planet is subject to type I migration, since the disc is hardly
perturbed. To investigate this further, we calculated the gap open criterion by
\cite{2006Icar..181..587C}
\begin{equation}\label{eq:gap_opening}
    P = \frac{h}{q^{1/3}} + \frac{50\alpha h^2}{q} \leq 1 \,.
\end{equation}
For our range of planet-to-binary mass ratios and our fixed disc parameters
($h=0.05$, $\alpha=0.01$) the gap open criterion is not met, confirming the type
I migration regime. Even very high-mass planets cannot open a full gap ($P
\approx 4$); they only produce a small dip in the density profile of the disc.
This can be seen in Fig.~\ref{img:k38_sigma_prof_gap} for the case of
Kepler-38.

Our findings are in good agreement with \citet{2014A&A...564A..72K} who also did
not find a dependence of the final planet position on the initial position.
However, in their simulations, the planet migrated further in and stopped at $\ap
= 0.436\,\mathrm{au}$, which is very close to the observed location. However,
this extended inward motion was an artefact of the incorrect location of the inner
boundary of the grid for which an overly large inner radius was chosen.

Since the initial location of the planet does not impact its final stopping position,
we always place the planet for Kepler-38 and the other systems slightly inside
the peak density to save some computational time. The azimuthally averaged
density profiles of all systems after \num{10000} binary orbits together with
the initial planet positions can be seen in Fig.~\ref{img:sigma_prof_init_pos}.

\begin{figure}
    \centering
    \resizebox{\hsize}{!}{\includegraphics{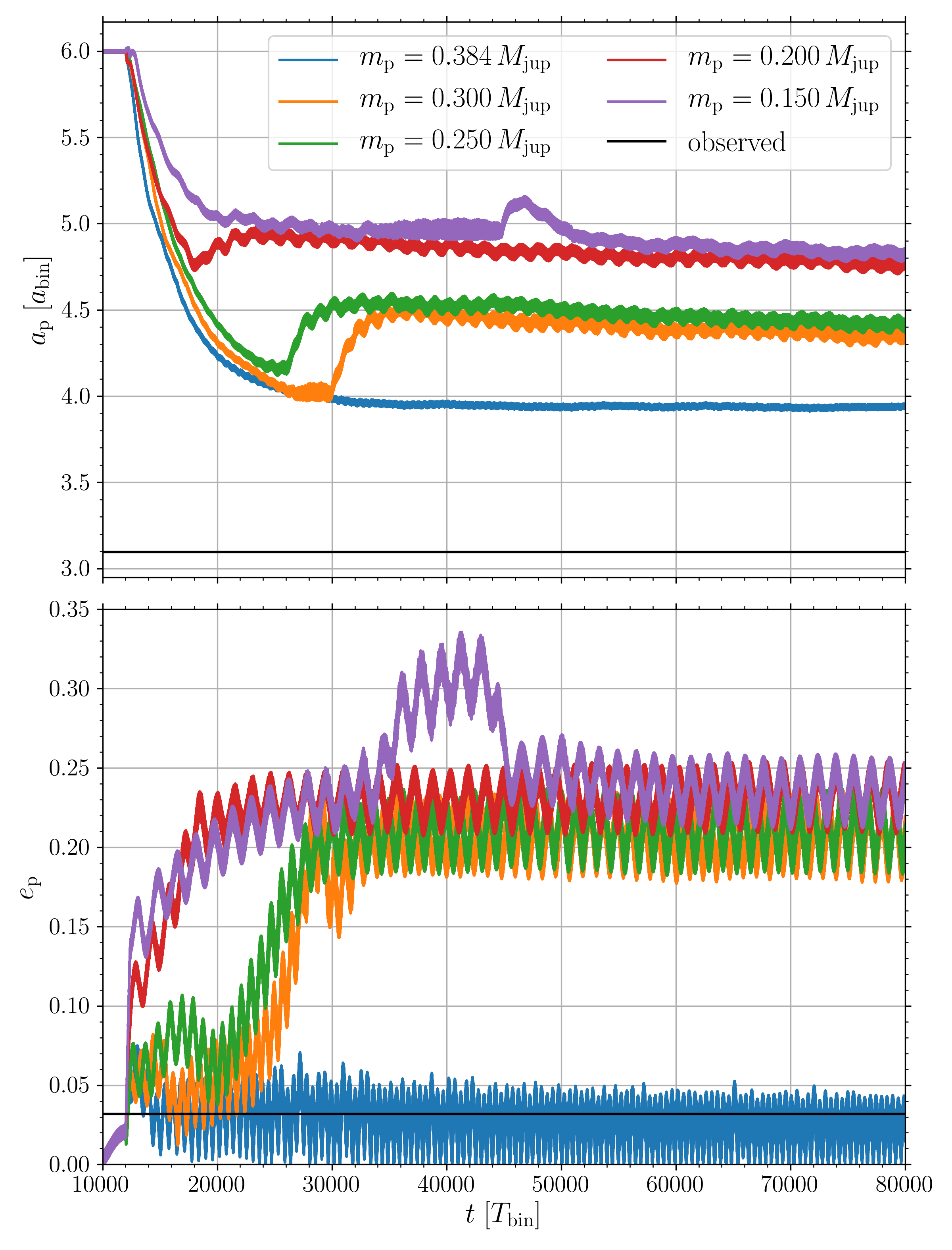}}
    \caption{Orbital elements of migrating planets in a circumbinary disc around
    the Kepler-38 system with different masses. \emph{Top:} Semi-major axis of
    the planet. \emph{Bottom} Eccentricity of the planet. The black lines show
    the observed values, for the eccentricity only an upper limit is known. The
    upper mass limit for Kepler-38b is $m_\mathrm{p} = 0.384\,M_\mathrm{jup}$.}
    \label{img:k38_orbital_elements}
\end{figure}

\begin{figure}
    \centering
    \resizebox{\hsize}{!}{\includegraphics{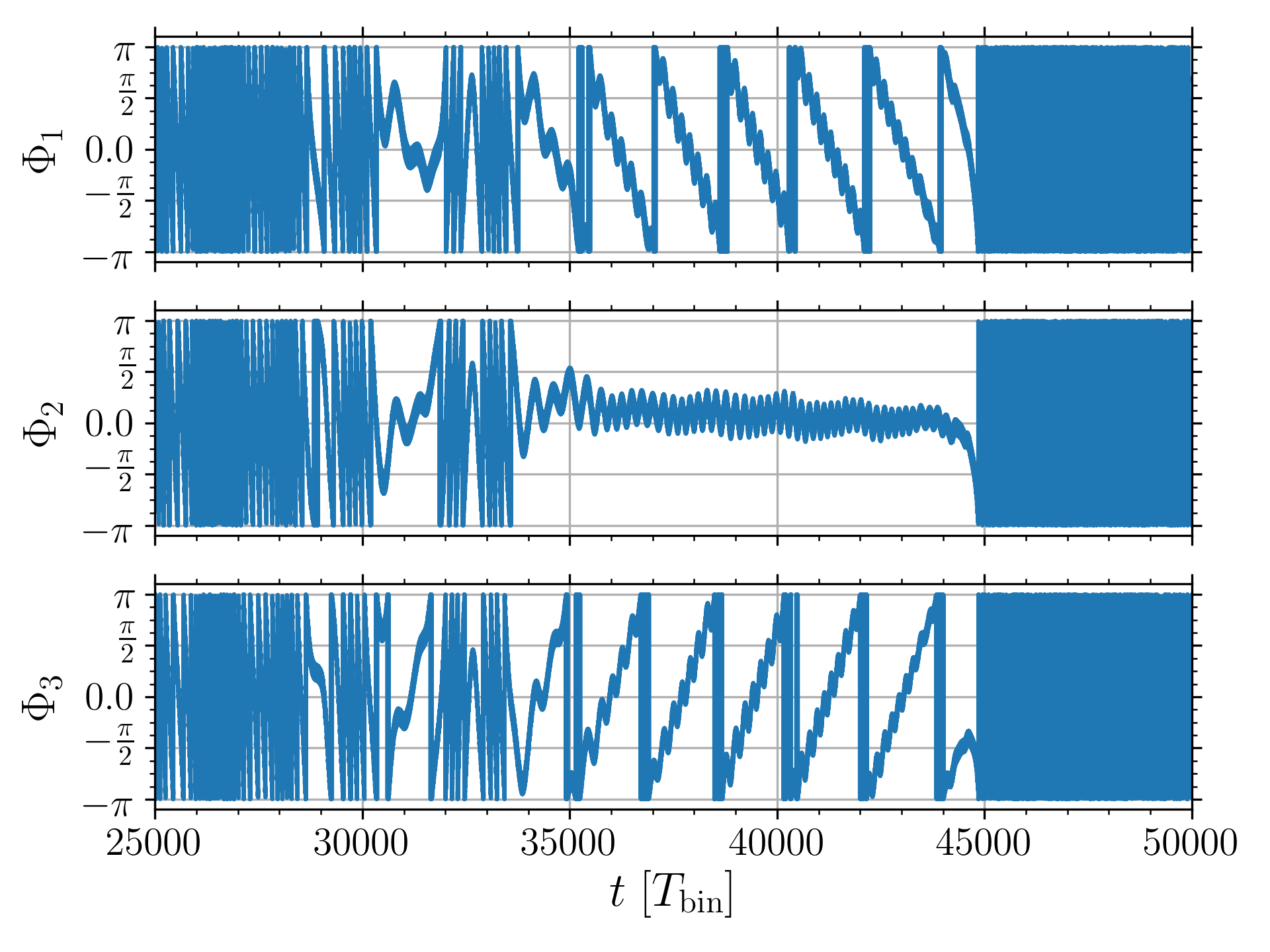}}
    \caption{Three out of eleven resonant angles for the $m_\mathrm{p} =
    0.150\,M_\mathrm{jup}$ case. While the resonant angles $\Phi_1$ and $\Phi_3$
    cover the full range from $0$ to $2\pi$ the resonant angle $\Phi_2$ librates
    around zero during the resonant phase of the planet 
    (from $t=\num{34000}\,T_\mathrm{bin}$ to $t=\num{45000}\,T_\mathrm{bin}$).
    The resonant angles $\Phi_3$ to $\Phi_{11}$ also cover the full range from
    $0$ to $2\pi$ but are not shown for clarity.}
    \label{img:k38_resonant_angles}
\end{figure}
%
\subsubsection{Variation of planet mass}\label{sssec:var_planet_mass}
In this section, we investigate the influence of the planet mass on the migration
process. In Fig.~\ref{img:k38_orbital_elements}, we display the time evolution of
the semi-major axis (top panel) and the eccentricity (bottom panel) of planets
with different masses which range here from $m_\mathrm{p}=0.150\,M_\mathrm{jup}$ to
$m_\mathrm{p}=0.384\,M_\mathrm{jup}$, the upper limit of the observed planet.  One can
see that, depending on the mass of the planet, two eccentricity states exist and
that the migration path of the planet through the disc differs for those two
states. The heaviest planet (with  $m_\mathrm{p}=0.384\,M_\mathrm{jup}$) migrates
smoothly to its final position and its eccentricity remains very low.  The
lighter planets start migrating inwards until they reach a turning point ( the lighter the planet is,  the earlier this
point is reached). From there on, the
planet undergoes a fast and short period of outward migration, during which the
eccentricity of the planet is increased to $\approx 0.20 - 0.22$. This increase
in eccentricity does not depend strongly on the planet mass. After the
eccentricity is increased, the planets migrate towards the binary very slowly,
compared to the first inward migration phase.

\begin{figure*}
    \centering
    \includegraphics[width=17cm]{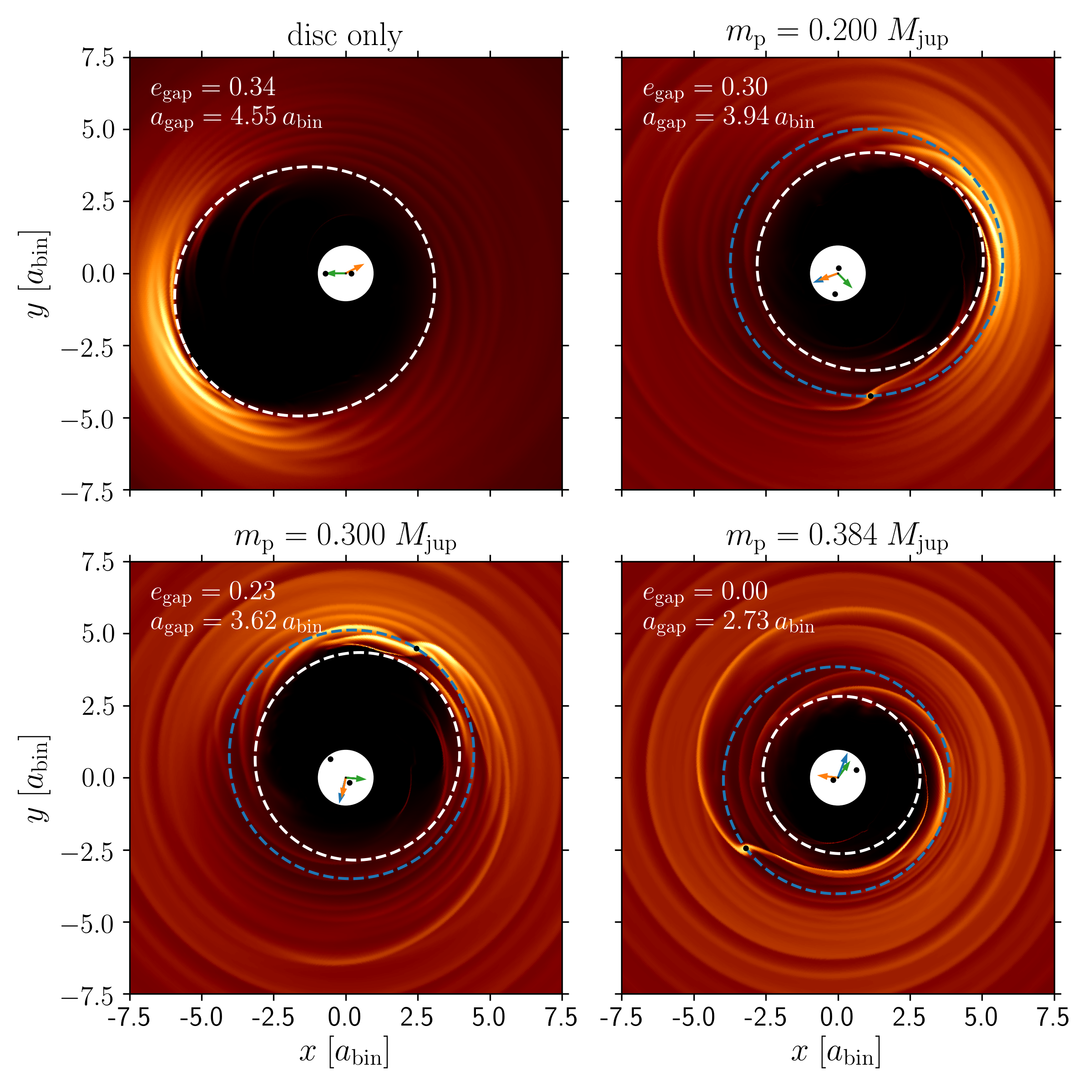}
    \caption{Structure of the inner disc for the Kepler-38 system. The surface
    density is colour-coded, since the absolute density values are not important
    for the disc structure we omit the density scale for clarity (brighter
    colours mean higher surface densities). The top-left panel shows the
    circumbinary disc after \num{10000} binary orbits before the planet
    is inserted. The remaining three panels show the disc structure with an
    embedded planet of varying mass after \num{76500} binary orbits. The binary
    components and the planets are marked with black dots. The
    direction of pericentre of the secondary (green), the planet (blue) and the disc (orange)
    are marked by the arrows. Additional fitted ellipses to the central
    cavity are plotted in dashed white lines and the actual orbit of the
    planet is shown as dashed blue line. Using the fitted ellipses, the size and
    eccentricity of the disc gap, shown in the upper left edge of each panel, are calculated at the same
    time. As we saw in
    Fig.~\ref{img:disc_properties} these quantities show small variations in time around some
    average value.}
    \label{img:2d_sigma_planet}
\end{figure*}
Before discussing the origin of these different migration scenarios, we focus
briefly on a special behaviour of the lightest planet with mass $m_\mathrm{p} =
0.150\,M_\mathrm{jup}$. At around $t \approx \num{34000}\,T_\mathrm{bin}$ into
the evolution, one notices an additional increase in eccentricity of the planet
while the migration seems to stall at around $\ap = 4.93\,a_\mathrm{bin}$; see
purple curves in Fig.~\ref{img:k38_orbital_elements}.  This phase of constant
semi-major axis and increase of eccentricity lasts until \num{45000} binary
orbits.  During this phase, we find for the quotient of the planet's period to
the binary period, $T_\mathrm{p}/T_\mathrm{bin} \approx 11.2$, suggesting a
possible capture into the 11:1 resonance between the planet and the binary as
the reason for this behaviour of the planet.  The position of this 11:1
resonance lies at $a_{11:1} = 4.93\,a_\mathrm{bin}$.  To check for this idea, we
calculated the $k$ resonance angles, which are for a general $p$:$q$ resonance
with $p > q$ given by
\begin{equation}\label{eq:resonant_angles}
    \Phi_k = p \lambda_\mathrm{p} - q \lambda_\mathrm{bin} - p \varpi_\mathrm{p} + q
    \varpi_\mathrm{bin} + k (\varpi_\mathrm{p} - \varpi_\mathrm{bin})
,\end{equation}
with $q \leq k \leq p$. Here, $\lambda_\mathrm{p}$ and $\lambda_\mathrm{bin}$
are the mean longitudes of the planet and the binary and $\varpi_\mathrm{p}$ and
$\varpi_\mathrm{bin}$ the longitudes of periapse. The planet and the binary are
in a $p$:$q$ resonance when at least one resonant angle $\Phi_k$ does not cover
the full range from $0$ to $2\pi$
\citep{2002MNRAS.333L..26N,2014A&A...564A..72K}. In our case, $\Phi_2$ librates
around zero and all other angles cover the full range.
Figure~\ref{img:k38_resonant_angles} shows the first three resonant angles as a
selection to illustrate the behaviour of the librating angle and angles which
cover the full range. This confirms that the planet is indeed captured in a 11:1
resonance with the binary. After \num{45000} binary orbits, the planet is kicked
out of this resonance, its semi-major axis increases shortly and its
eccentricity decrease to the value before the resonant capture occurred. After
this short resonant capture, the planet resumes its slow inward migration, similarly to  the
other planets.  As seen in Fig.~\ref{img:k38_resonant_angles}, previous to
this extended phase of being engaged in the 11:1 resonance, there was a brief
capture into the same resonance at $t \approx \num{31000}\,T_\mathrm{bin}$.

Coming back to the interaction of the planets with the disc, the impact an
embedded planet has on the ambient disc with constant aspect ratio and
constant turbulent viscosity depends on its mass. For the most
massive planet, we found that the inner disc became more circular with a reduced
size of the inner gap. This allowed the planet to migrate further inward towards
the binary. Lighter planets cannot alter the disc structure in such a strong way
due to the reduced gravitational impact.  The change of the inner disc structure
in the presence of a planet is illustrated in Fig.~\ref{img:2d_sigma_planet}.
This figure shows the 2D surface density for three different planet masses as well as
the disc structure prior to inserting the planet (top left panel). The white
dashed lines show the fitted ellipses to the inner cavity. One can immediately
see how the size and eccentricity of the gap decrease with increasing planet
mass. Another observation is that the final position of the planet is always
close to the edge of the gap. The dashed blue lines show the orbit of the
planet, which lies in all cases just outside the fitted ellipse.  Going from the
upper left panel to the lower right panel, we see the clear trend that
eccentricity and semi-major axis of the gap is decreasing as the planet mass is
increased.

\begin{figure}
    \centering
    \resizebox{\hsize}{!}{\includegraphics{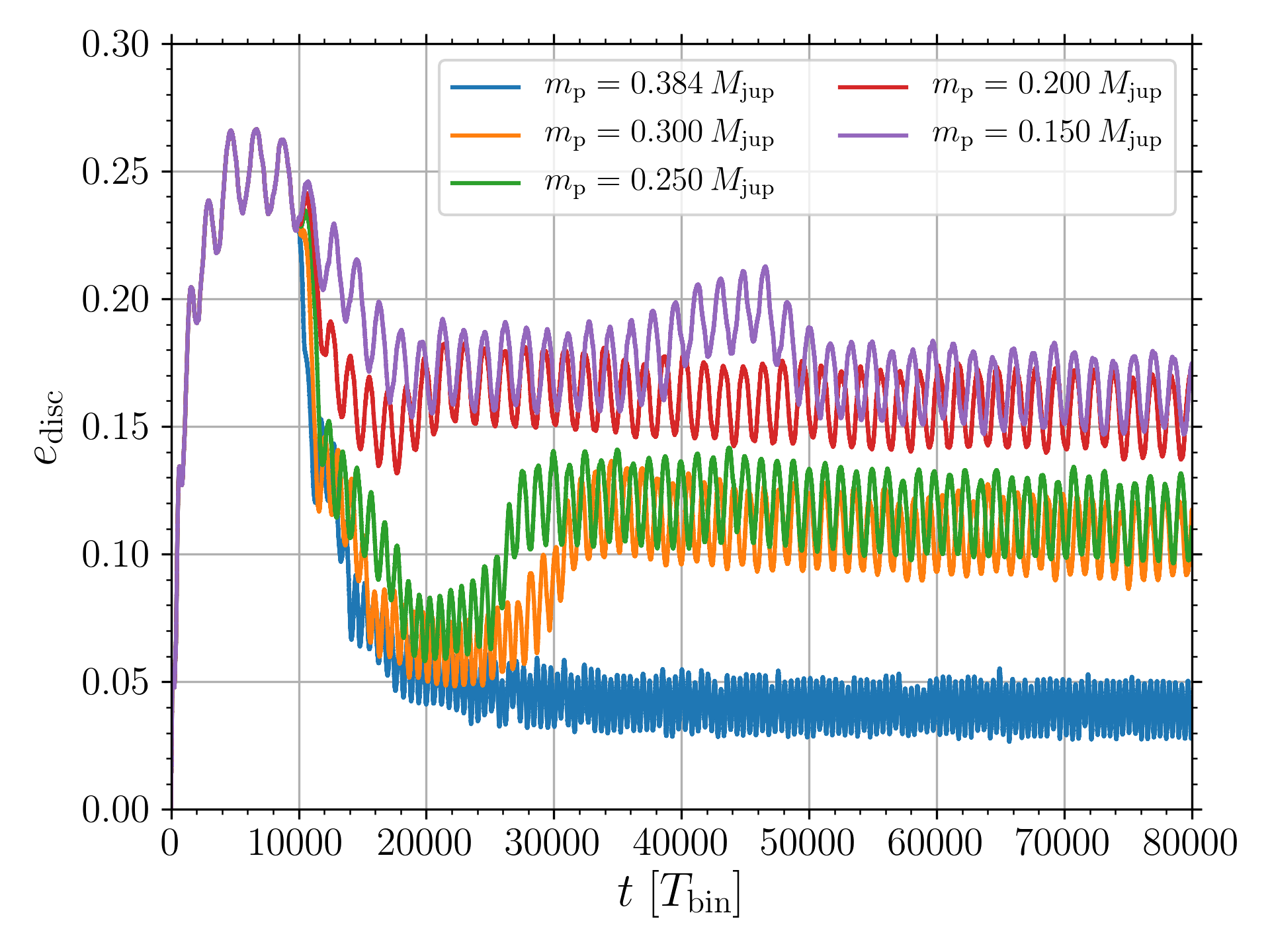}}
    \caption{Time evolution of the inner disc eccentricity for Kepler-38.
    Planets of different
    mass were introduced into the simulations at \num{10000} binary orbits on
    circular orbits
    and allowed to migrate through the disc after \num{12000} binary orbits.}
    \label{img:k38_disc_ecc_with_planet}
\end{figure}
\begin{figure}
    \centering
    \resizebox{\hsize}{!}{\includegraphics{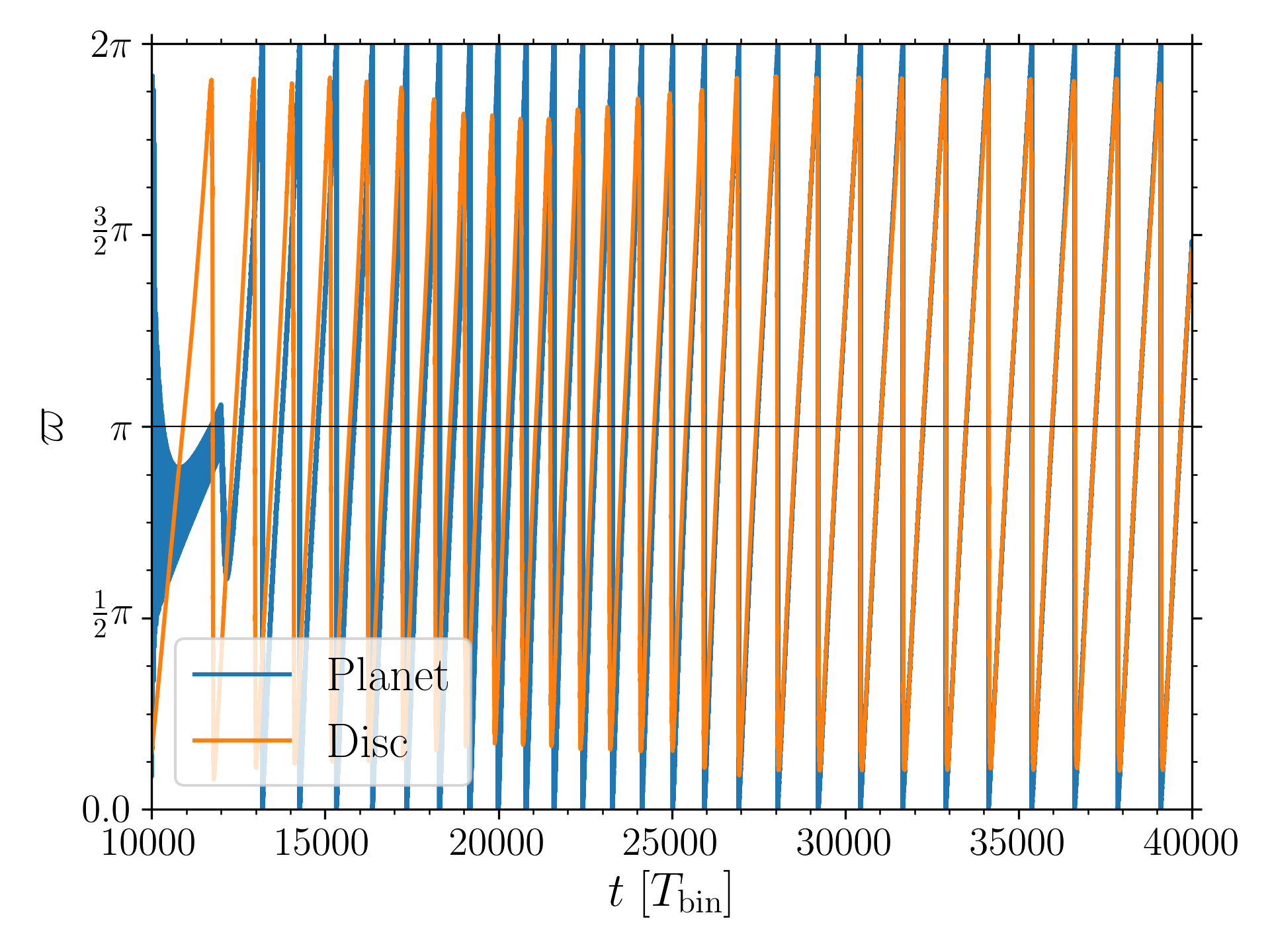}}
    \caption{Time evolution of the pericentres of a planet with mass
    $m_\mathrm{p} =
    0.250\,M_\mathrm{jup}$ (blue) and the disc (orange) for Kepler-38.
    Until $t = \num{12000}\,T_\mathrm{bin}$ the planet is on a fixed circular orbit and
    therefore the pericentre is not well defined. After switching on the 
    back-reaction onto the planet, its orbit soon becomes aligned with the inner
    precessing cavity of the disc.}
    \label{img:disc_planet_peri}
\end{figure}

A more quantitative image of this behaviour can be seen in
Fig.~\ref{img:k38_disc_ecc_with_planet}, which shows the time evolution of the
inner disc eccentricity. The disc eccentricity and the disc pericentre (in
Fig.~\ref{img:disc_planet_peri}) are calculated according to \citet[Sect.
2.5]{2017A&A...604A.102T} within a limited radial interval $[R_1,R_2]$, where
$R_1 = R_\mathrm{min}$ and $R_2$ is chosen differently for each system such that
it lies at a point slightly outside the density maximum. For the five systems
considered in this paper, we have chosen $R_2 = \{7.0, 12.0, 7.0, 7.0,
9.0\}\,a_\mathrm{bin}$ (compare to Fig.~\ref{img:sigma_prof_init_pos}). For the
first \num{10000} binary orbits, no planet is present in the simulations and the
disc eccentricity rises to about $e_\mathrm{disc} \approx 0.25$.  After
\num{10000} binary orbits, the planet is introduced into the simulation, and
after \num{12000} binary orbits, the planet is allowed to migrate through the
disc. At the moment the planet is embedded, the disc's eccentricity decreases;
the heavier the planet is, the more abrupt the decrease. For the light planets,
except the planet with mass $m_\mathrm{p}=0.150\,M_\mathrm{jup}$, the disc
\enquote{pushes} back, increasing its eccentricity again. From this point, the
inner disc eccentricity stays more or less constant. This push-back phase
coincides with the outward migration phase of the planet (see
Fig.~\ref{img:k38_orbital_elements}).  Interestingly, this increase of the disc
eccentricity only increases the planet's semi-major axis but not the
eccentricity of the planet. In the case of the massive planet, the disc is not
able to \enquote{push back}, as it is not massive enough, and the disc
eccentricity continues to decrease smoothly to a final value of $e_\mathrm{disc}
= 0.043$.
As indicated by the arrows in Fig.~\ref{img:2d_sigma_planet}, which mark the
pericentre of the secondary (green), the planet (blue) and the disc (orange),
the pericentre of the disc and the planet are aligned in the case of the light
planets. For the massive planet, no such alignment is observed; the similar
directions of disc and planet pericentre at this snapshot are coincidental.
Also, the definition of a unique pericentre becomes difficult in the case of
massive planets, because the planet orbit and the disc gap have an eccentricity
close to zero. The time evolution of the directions of pericentre for disc and
planet is shown in Fig.~\ref{img:disc_planet_peri} for the $m_\mathrm{p} =
0.250\,M_\mathrm{jup}$ case. The orbital alignment between disc and planet is
achieved during the first \num{10000} binary orbits after releasing the planet.

\citet{2015A&A...581A..20K} and \citet{2017MNRAS.469.4504M} also found this
alignment of planet orbit and inner disc cavity for the Kepler-34 system.

\subsubsection{Variation of disc mass}
As we have seen in the previous section, the planet can, depending on its mass,
modify the disc structure. To investigate now the influence of the disc mass on
this behaviour, we set up simulations with a different disc mass before
introducing the planet into the disc. In all our models, we started with an
initial disc mass of $M_\mathrm{disc} = 0.01\,M_\mathrm{bin}$. Since material
can leave the computational domain through the inner open boundary, the disc will
lose mass over time. After \num{10000} binary orbits, before adding the planet
to the simulation, the disc in the standard model has a mass of $M_{10000}
\equiv M_\mathrm{disc}(t=\num{10000}\,T_\mathrm{bin}) =
0.008967\,M_\mathrm{bin}$.  For two different planet masses, we increased or
decreased the disc mass and followed the evolution of the embedded planets.  Due
to the local isothermal assumption, the neglect of the disc self-gravity and the
independence of the potential of the surface density (the back-reaction of the
disc onto the binary is not yet switched on, this means the disc terms in
eq.~\eqref{eq:acc_com} and \eqref{eq:eq_of_motion} are neglected), the mass of the
disc can be rescaled without changing its dynamics. The set of models using this
disc mass rescaling is summarised in Table~\ref{tab:planet_disc_q}.
\begin{table}
    \caption{Planet mass $m_\mathrm{p}$, disc mass $M_\mathrm{disc}$ and the
    ratio $q = m_\mathrm{p}/M_\mathrm{disc}$ for models using different disc masses.  The
    quantity $M_{10000} = 0.008967\,M_\mathrm{bin}$ refers to the disc mass
    after \num{10000} orbits of the standard model for Kepler-38 without an embedded
    planet.}
    \label{tab:planet_disc_q}
    \centering
    \begin{tabular}{c c c}
        \hline\hline
        \noalign{\smallskip}
        $m_\mathrm{p}\;[M_\mathrm{jup}]$ &
        $M_\mathrm{disc}\;[M_\mathrm{bin}]$ &
        $q$ \\
        \noalign{\smallskip}
        \hline
        \noalign{\smallskip}
        0.384 & $M_{10000}$ & 0.03388 \\
        \\
        0.200 & $M_{10000}$ & 0.01765 \\
        \noalign{\smallskip}
        \hline
        \noalign{\smallskip}
        0.384 & $ 3 M_{10000}$  & 0.01142 \\
        \\
        0.200 & $ \frac{1}{2} M_{10000}$ & 0.03568 \\
        \noalign{\smallskip}
        \hline
    \end{tabular}
\end{table}
The top part of Table~\ref{tab:planet_disc_q} gives an overview of the
planet-disc mass ratio of two models, using the standard disc mass. As seen
in the previous section, the planet with mass $m_\mathrm{p} = 0.384\,M_\mathrm{jup}$,
which was the most massive planet, could alter the disc structure,
whereas a planet with mass $m_\mathrm{p} = 0.200\,M_\mathrm{jup}$ could not.
The bottom part of Table~\ref{tab:planet_disc_q}
shows the planet-disc mass ratio for models with increased or decreased disc
mass. For the $m_\mathrm{p} = 0.384\,M_\mathrm{jup}$ model, we increased the disc mass by
a factor of three so that the planet-disc mass ratio is now comparable to the
$m_\mathrm{p} = 0.200\,M_\mathrm{jup}$ case with standard disc mass. Therefore, we expect
that now the massive planet cannot alter the disc structure and will end up in a
state with high eccentricity. In the case where $m_\mathrm{p} = 0.200\,M_\mathrm{jup}$, we
decreased the disc mass by a factor of one half to increase the disc-planet mass
ration to the regime where the planet can alter the disc. In this case, therefore, the
planet should migrate smoothly inwards and end up in a state with very low
eccentricity.

\begin{figure}
    \centering
    \resizebox{\hsize}{!}{\includegraphics{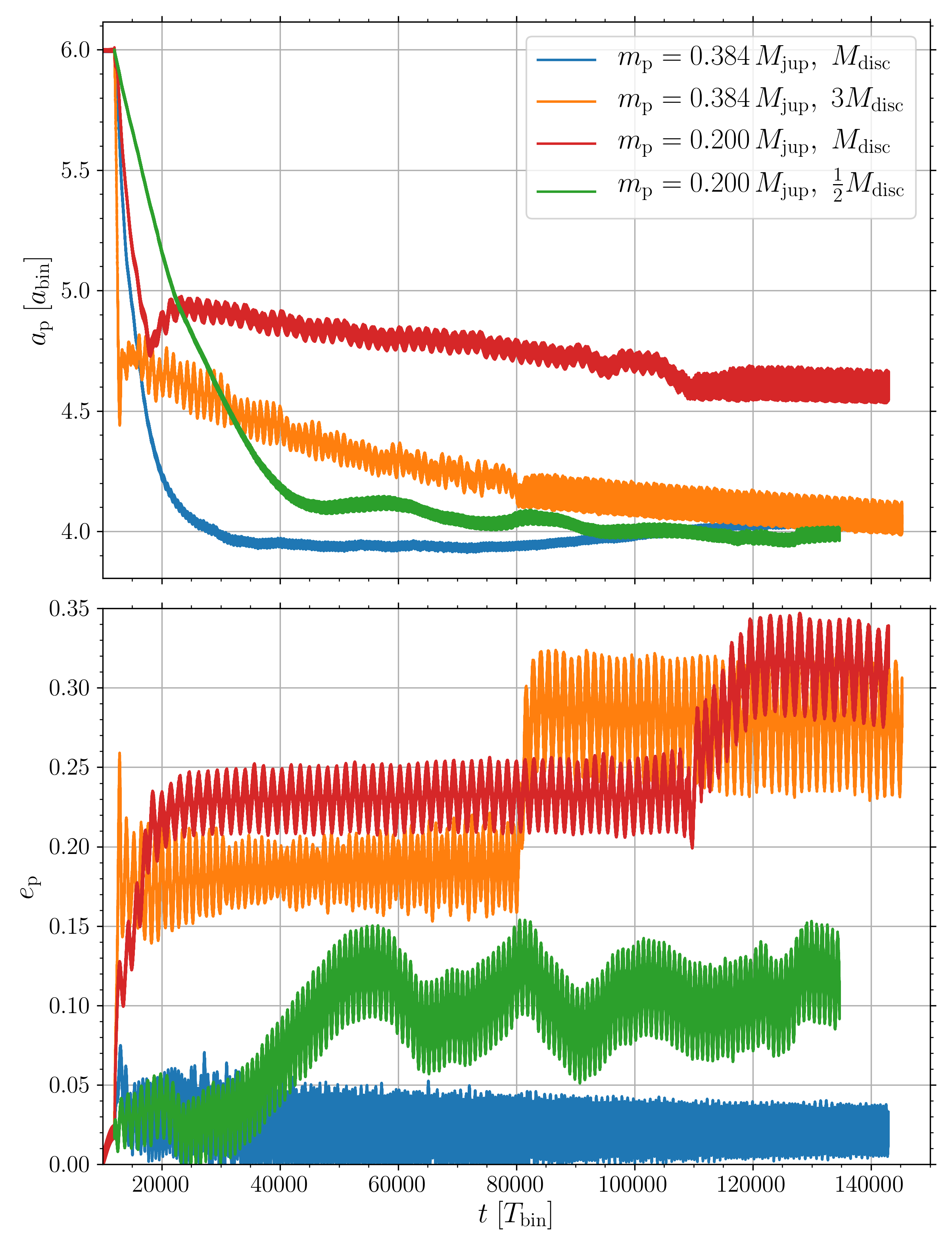}}
    \caption{Time evolution of semi-major axis (top) and eccentricity (bottom)
    of planets with different mass in discs of different mass, for the Kepler-38
    system.}
    \label{img:k38_orbital_elements_disc_mass}
\end{figure}
The results of these simulations are summarised in
Fig.~\ref{img:k38_orbital_elements_disc_mass}.  The colours for the
simulations with the standard disc mass are identical to
Fig.~\ref{img:k38_orbital_elements}.  The migration speed of the $m_\mathrm{p} =
0.384\,M_\mathrm{jup}$ planet in the disc with triple mass is very rapid
compared to the standard disc. Since the migration speed is directly
proportional to the disc mass, this behaviour is expected.  This fast inward
migration period lasts for approximately \num{552} binary orbits, during which
the eccentricity of the planet is increased to $\ep = 0.184$.  Following that
phase, the planet migrates again outward for a very short period of time, as
already seen for low-mass planets in the standard disc. After that short period
of outward migration, the planet resumes its inward migration but with a very low
speed compared to the initial inward migration period.  This is probably
due to the high eccentricity of the planet. At around \num{80000} binary orbits,
the planet is captured in a 9:1 resonance with the binary, which increases its
eccentricity further to a final value of $\ep = 0.273$. The capture in resonance
is confirmed by analysing the resonant angles; see eq.~\eqref{eq:resonant_angles}.
Again, $\Phi_2$ librates around zero, while all the other angles circulate and
cover the full range. Although the planet is in resonance with the binary its
semi-major axis still decreases slowly. This is due to the fact that the binary
separation decreases faster than in the standard case due to the three times increased disc mass.  This time, the planet's orbit is aligned with the
precessing inner gap of the disc. 

The $m_\mathrm{p} = 0.2\,M_\mathrm{jup}$ planet in the standard-disc-mass case is
captured in a 8:1 resonance with the binary after $t =
\num{110000}\,T_\mathrm{bin}$.
For the reduced-disc-mass case (last row in Tab.~\ref{tab:planet_disc_q}), the
$m_\mathrm{p} = 0.2\,M_\mathrm{jup}$ planet migrates slowly through the disc with no
turning point. At the beginning, the eccentricity of the planet is close to zero,
as expected, because due to the reduced disc mass the planet is now able to
circularise the disc.  Additionally, the orbit of the planet is not aligned with
the inner gap.  At around \num{30000} binary orbits, the eccentricity increases
and starts to fluctuate around $\ep = 0.10$ on a long timescale.

After having described the main features of planet migration in circumplanetary
discs for the Kepler-38 system, we turn now to the other four systems. As
before, we run the disc with initially $10^{-2}$ $M_\mathrm{bin}$ without a
planet for $\num{10000}\,T_\mathrm{bin}$ and then embed planets of different
masses. For the first $\num{2000}\,T_\mathrm{bin}$ the planet is fixed at its orbit,
then released and evolved for about $\num{70000}\, T_\mathrm{bin}$.  The final
orbital parameters for the planets with the observed mass can be found in
Table~\ref{tab:orbital_elements}, together with the observed values. 

\begin{figure}
    \centering
    \resizebox{\hsize}{!}{\includegraphics{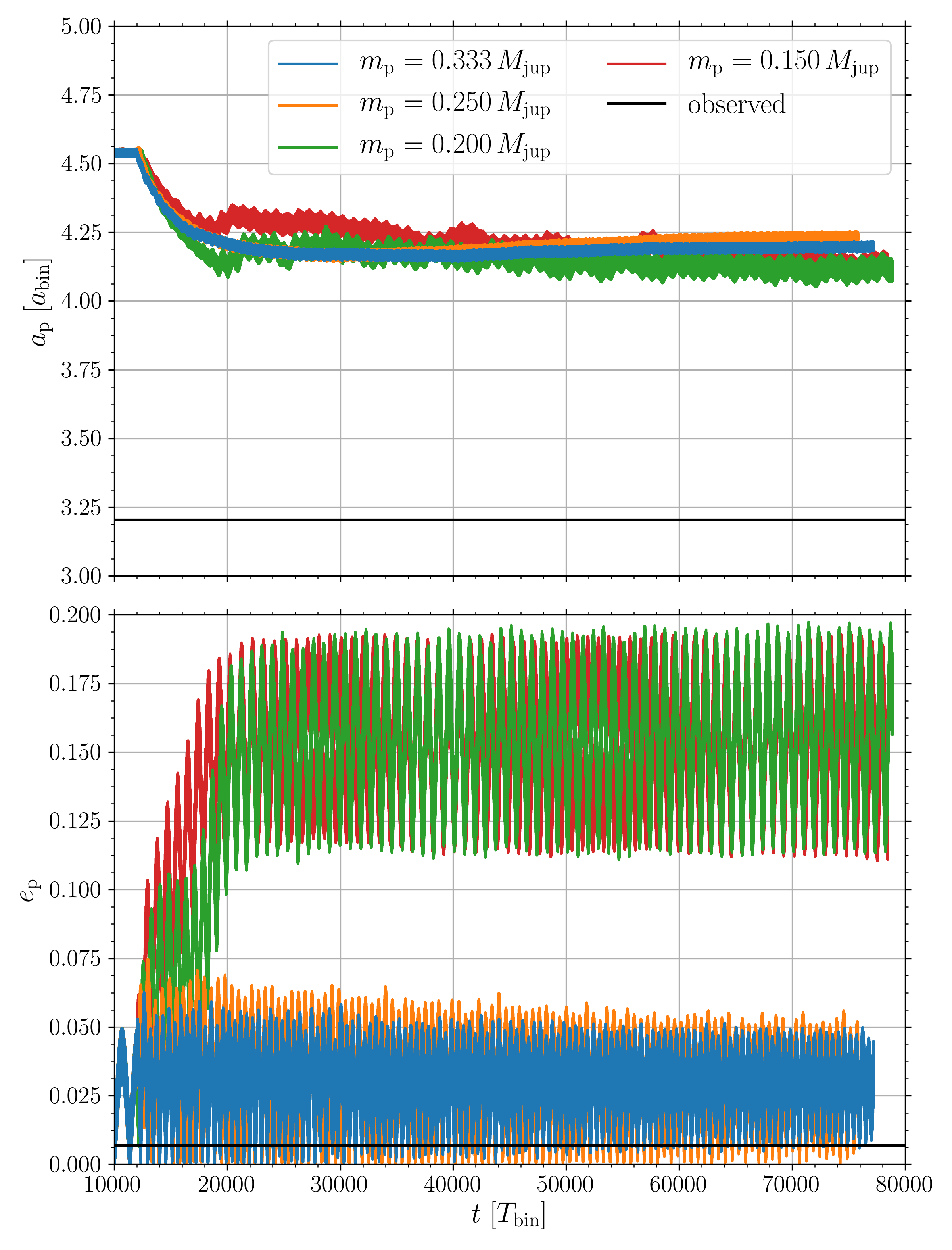}}
    \caption{Orbital elements of migrating planets in a circumbinary disc around
    the Kepler-16 system with different masses. \emph{Top:} Semi-major axis of
    the planet. \emph{Bottom} Eccentricity of the planet. The black lines show
    the observed values. The observed mass of Kepler-16b is 
    $m_\mathrm{p} = 0.333\,M_\mathrm{jup}$.}
    \label{img:k16_orbital_elements}
\end{figure}
\subsection{Kepler-16}
To study the evolution of planets in the Kepler-16 system, we performed
simulations with different planet masses starting from the observed mass of
$m_\mathrm{p}
= 0.333\,M_\mathrm{jup}$ down to a mass of $m_\mathrm{p} = 0.150\,M_\mathrm{jup}$.
Figure~\ref{img:k16_orbital_elements} shows the time evolution of the semi-major
axis (top) and eccentricity (bottom) of the different planets.  As seen in the
eccentricity evolution, there are two distinct states, similar to the Kepler-38
case. The eccentricity of the more massive planets is not excited and remains
close to zero, oscillating around $\ep = 0.029$, whereas the eccentricity of the
lighter planets is increased during the first \num{8000} binary orbits, after
releasing the planets, to a mean
value of $\ep = 0.15$. These two states also differ in their alignment of the
orbits with the inner gap of the disc. The lighter planets with high
eccentricity end up in a state with aligned orbits, whereas no such alignment can
be observed for the heavier planets.  In the semi-major axis evolution, these two
states are not as clearly visible as in the Kepler-38 system. Nonetheless, a
difference between the massive and light planets can be observed. Heavy planets
migrate smoothly through the disc, whereas the semi-major axis evolution of the
lighter planets shows more noise.  The massive planets start to migrate inwards
smoothly, as expected from the Kepler-38 case. This phase smoothly transitions
to a slow outward migrating phase, as seen also in long time simulations of
Kepler-38 (see blue curve in Fig.~\ref{img:k38_orbital_elements_disc_mass}). The
lighter planets have also a period of outward migration, but this period is very
short and followed by a second period of inward migration, as in the Kepler-38
case. The migration speed during this second inward period is far lower than
during the first phase. The light planets show the same general behaviour as in
the Kepler-38 case and they even migrate further in
than their heavier counterparts.  One reason for this could be that Kepler-16
has the smallest gap with the lowest eccentricity and therefore the
circularising effect of heavier planets is not as important as in the other
systems. This also implies that the increase of eccentricity is connected to the
alignment of the planets orbit with the disc gap.

\begin{table}
    \caption{Final orbital parameters of the embedded planets using the observed
    $m_\mathrm{p}$ for the five systems investigated, calculated using the
    \textsc{Pluto} simulations. The observed data are displayed
    also, for comparison.}
    \label{tab:orbital_elements}
    \centering
    \begin{tabular}{c c c c c c}
        \hline\hline
        \noalign{\smallskip}
        System & 
        $\ap\;[a_\mathrm{bin}]$ &
        $a_\mathrm{obs}\;[a_\mathrm{bin}]$ & 
        $\ep$ &
        $e_\mathrm{obs}$\\
        \noalign{\smallskip}
        \hline
        \noalign{\smallskip}
        Kepler-16 & 4.18 & 3.20 & 0.03 & 0.00685 \\ 
        \\
        Kepler-34 & 7.77 & 4.74 & 0.31 & 0.182 \\
        \\
        Kepler-35 & 4.47 & 3.35 & 0.12 & 0.042 \\
        \\
        Kepler-38 & 3.98 & 3.10 & 0.02 & 0.032 \\
        \\
        Kepler-413 & 5.53 & 3.55 & 0.30 & 0.1181 \\
        \noalign{\smallskip}
        \hline
    \end{tabular}
    \tablefoot{The semi-major axis ($\ap$) and eccentricity ($\ep$) of the
    planets are obtained through time averages over several thousand binary
    orbits at the end of the simulations.}
\end{table}

\begin{figure}
    \centering
    \resizebox{\hsize}{!}{\includegraphics{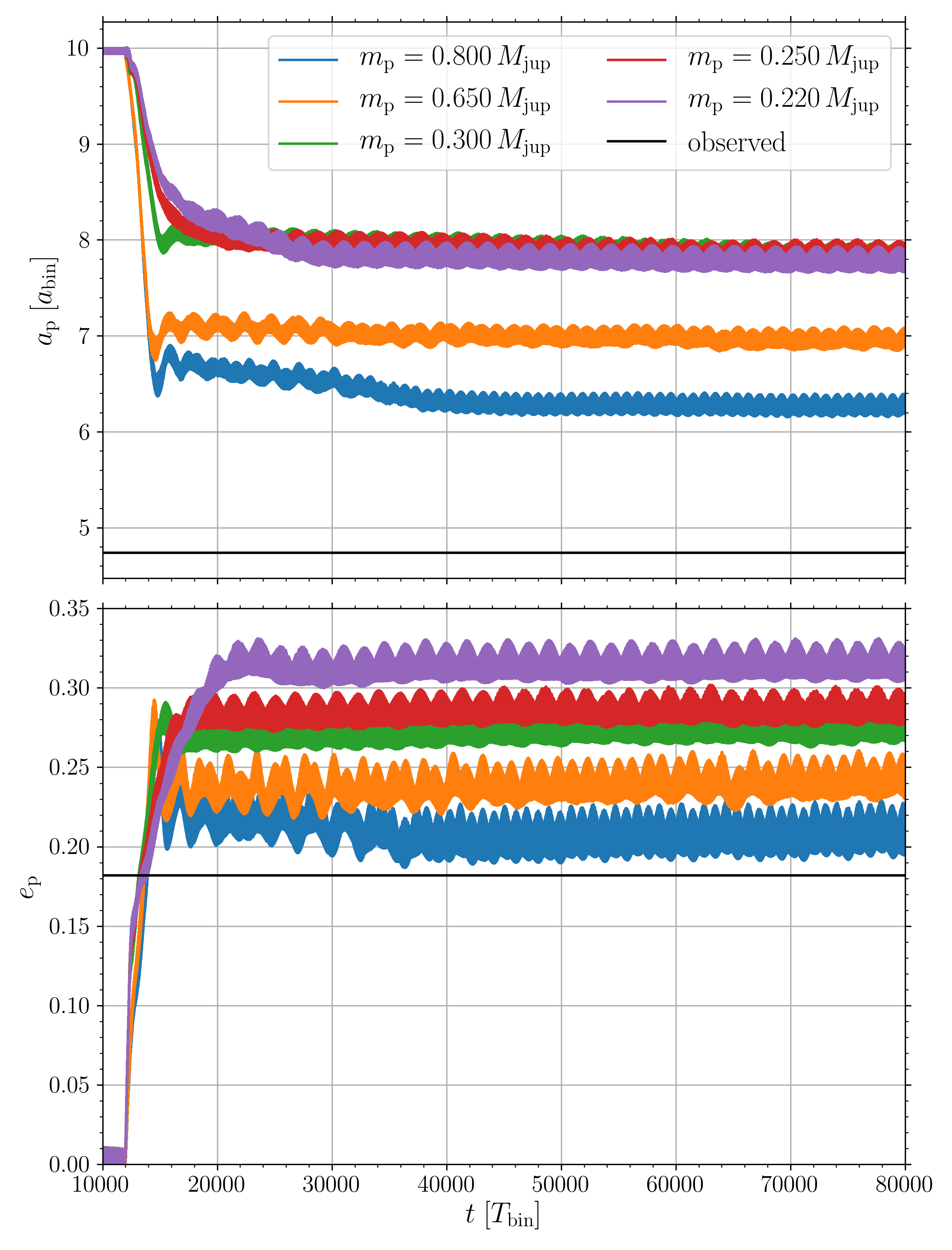}}
    \caption{Orbital elements of migrating planets in a circumbinary disc around
    the Kepler-34 system with different masses. \emph{Top:} Semi-major axis of
    the planet. \emph{Bottom} Eccentricity of the planet. The black lines show
    the observed values. The observed mass of Kepler-34b is 
    $m_\mathrm{p} = 0.220\,M_\mathrm{jup}$.}
    \label{img:k34_orbital_elements}
\end{figure}
\subsection{Kepler-34}
For the Kepler-34 system, we simulated planets with masses from the observed mass
of $m_\mathrm{p} = 0.220\,M_\mathrm{jup}$ up to $m_\mathrm{p} = 0.800\,M_\mathrm{jup}$.
Figure~\ref{img:k34_orbital_elements} gives an overview of the orbital elements
(semi-major axis and eccentricity) of the simulated planets. In all cases, we
observe nearly the same migration behaviour, first a fast inward migration
followed by a short period of outward migration. After that, the planets are
relatively soon (after around \num{40000} $T_\mathrm{bin}$) parked in their final
position.  In all cases, the final orbits of the planets are relatively eccentric,
with $\ep$  well above 0.2, and they are fully aligned with the eccentric gap of
the disc and precess at the same rate.  Although the massive planets can
circularise the inner cavity slightly, this effect is not strong enough, because
of the large and highly eccentric initial gap ($a_\mathrm{gap} =
6.99\,a_\mathrm{bin}, e_\mathrm{gap} = 0.4$).  For the planet with the observed
mass of $m_\mathrm{p} = 0.220\,M_\mathrm{jup}$ we find a final semi-major axis of $\ap =
7.77\,a_\mathrm{bin}$ and an eccentricity of $\ep = 0.31$ (see also
Table~\ref{tab:orbital_elements}). 

In their study of the planet in Kepler-34,
\citet{2015A&A...581A..20K} observed the same migration behaviour as we
find here: A fast inward migration to the edge of the inner cavity. They
find in their simulations a smaller semi-major axis of $\ap =
6.1\,a_\mathrm{bin}$ and a comparable eccentricity of $\ep = 0.3$. One reason for
the difference in the planets semi-major axis could be again the position of the
inner computational boundary.
\citet{2015A&A...581A..20K} used a value of $R_\mathrm{min} =
1.47\,a_\mathrm{bin}$, which is too large \citep{2017A&A...604A.102T}. Furthermore,
they use a uniformly spaced grid in the $R$-direction with only \num{256} cells.
Therefore, their resolution in the inner part of the computational domain is very
low, which could be another reason for the observed differences, but
they also found  an alignment of the planet's orbit with the
inner eccentric cavity. 

For Kepler-34, \citet{2017MNRAS.469.4504M} found for their one- and two-MMSN disc
models (our disc mass lies in between these two) a final semi-major axis of $\ap
\approx 8.7\,a_\mathrm{bin}$ and an eccentricity of $\ep = 0.25-0.3$, in
relatively good agreement with our results.  Furthermore, they also find an
alignment of the planet's orbit with the disc cavity. They mention this
alignment only in the case of a planetary core with very low mass, but since
they see no big difference between the full mass planet and the planetary core,
we assume that this alignment also happens in the case of the fully grown planet. 

\begin{figure}
    \centering
    \resizebox{\hsize}{!}{\includegraphics{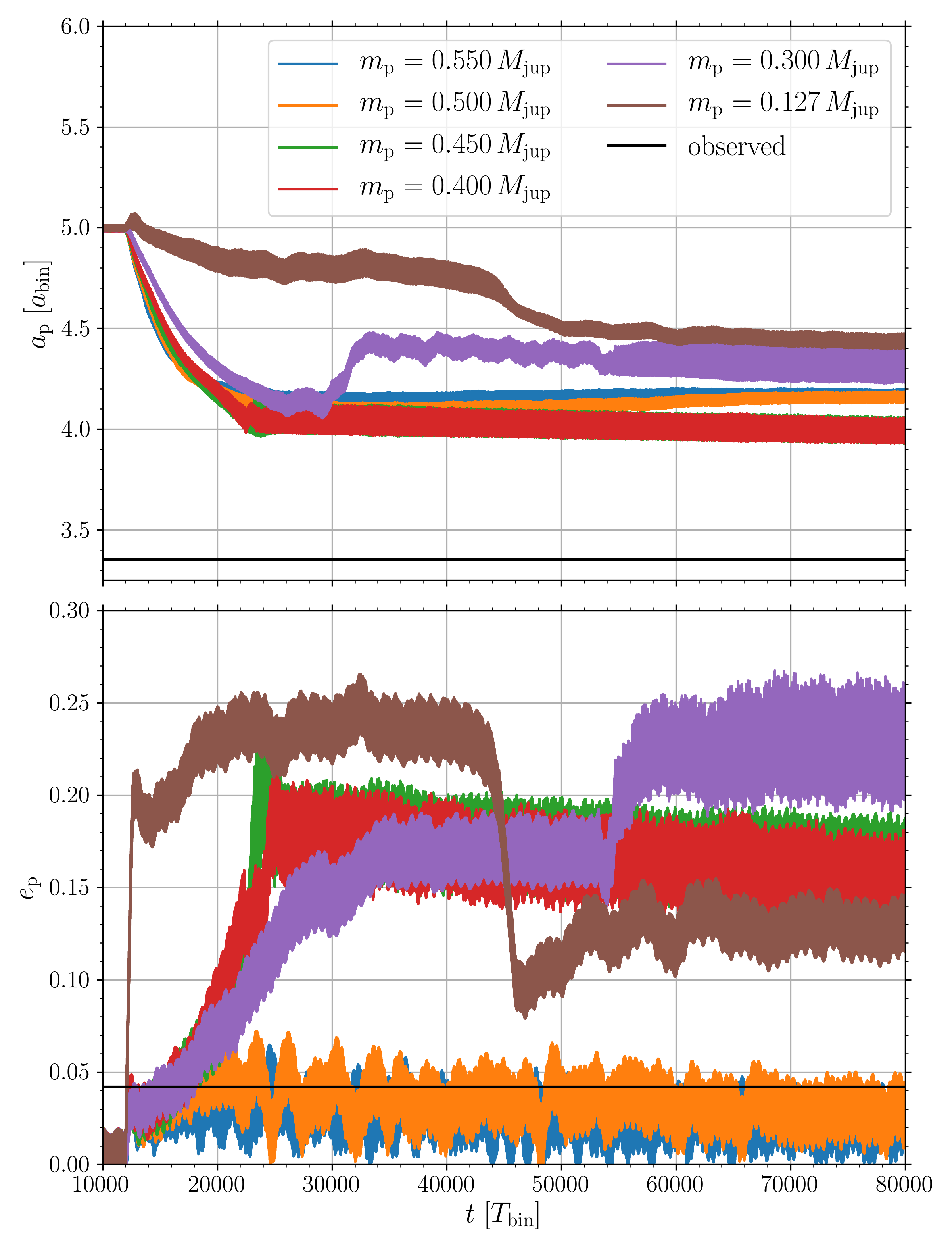}}
    \caption{Orbital elements of migrating planets in a circumbinary disc around
    the Kepler-35 system with different masses. \emph{Top:} Semi-major axis of
    the planet. \emph{Bottom} Eccentricity of the planet. The black lines show
    the observed values. The observed mass of Kepler-35b is 
    $m_\mathrm{p} = 0.127\,M_\mathrm{jup}$.}
    \label{img:k35_orbital_elements}
\end{figure}
\subsection{Kepler-35}\label{ssec:kepler35}
For Kepler-35, we studied the migration of planets starting from the observed
mass $m_\mathrm{p} = 0.127\,M_\mathrm{jup}$ up to $m_\mathrm{p} = 0.550\,M_\mathrm{jup}$.
Figure~\ref{img:k35_orbital_elements} shows results of simulations with the
different planet masses.  The two most massive planets with $m_\mathrm{p} =
0.550\,M_\mathrm{jup}$ and $m_\mathrm{p} = 0.500\,M_\mathrm{jup}$ show the typical
behaviour of massive planets, as already discussed for the other systems. Their
eccentricity remains low and they do not align their orbit with the precessing
inner gap of the disc. 

The eccentricity of planets with masses from $m_\mathrm{p} = 0.300\,M_\mathrm{jup}$ to
$m_\mathrm{p} = 0.450\,M_\mathrm{jup}$ become excited to a mean value of about $0.15$. The
planet with mass $m_\mathrm{p} = 0.300\,M_\mathrm{jup}$ ends up in a 9:1 resonance with
the binary after \num{55\,000} binary orbits, and therefore its eccentricity is increased
further. This planet also shows the general migration behaviour of lighter
planets, as seen for the other systems. This behaviour is not as clearly visible
for the planets with masses $m_\mathrm{p} = 0.400\,M_\mathrm{jup}$ and $m_\mathrm{p} =
0.450\,M_\mathrm{jup}$, since they are both captured in a 8:1 resonance with the
binary after $t \approx \num{23000}\,T_\mathrm{bin}$.

The planet with the observed mass of $m_\mathrm{p} = 0.127\,M_\mathrm{jup}$ shows a
behaviour not yet observed for the other systems. Its eccentricity is increased
fast to a value slightly below 0.25. After roughly \num{44000} binary orbits, its
eccentricity drops rapidly to a value below 0.15. This drop is also visible in
the semi-major axis of the planet.  A \textsc{Fargo3D} comparison simulation
with a planet of this mass shows the same behaviour, but at a later time.  The
fast drop in eccentricity happens there at roughly \num{59000} binary orbits.
Besides this time shift, the \textsc{Pluto} and \textsc{Fargo3D} curves for the
semi-major axis and the eccentricity have a very similarly evolution.
For both codes, the drop in eccentricity  (for $m_\mathrm{p} =
0.127\,M_\mathrm{jup}$) happens when the planet crosses $a_\mathrm{p} \approx
4.75\,a_\mathrm{bin}$.  Since in the case of \textsc{Fargo3D,} the planet
migrates slower, it needs more time to reach this point.
A reason for this behaviour could be the very low mass of the planet compared to
the disc in this case. Therefore, the influence of the disc on the planet is
very high, and the disc eccentricity drops simultaneously.
The final orbital parameters of the migrated planet are summarised
in Table~\ref{tab:orbital_elements}. 

\begin{figure}
    \centering
    \resizebox{\hsize}{!}{\includegraphics{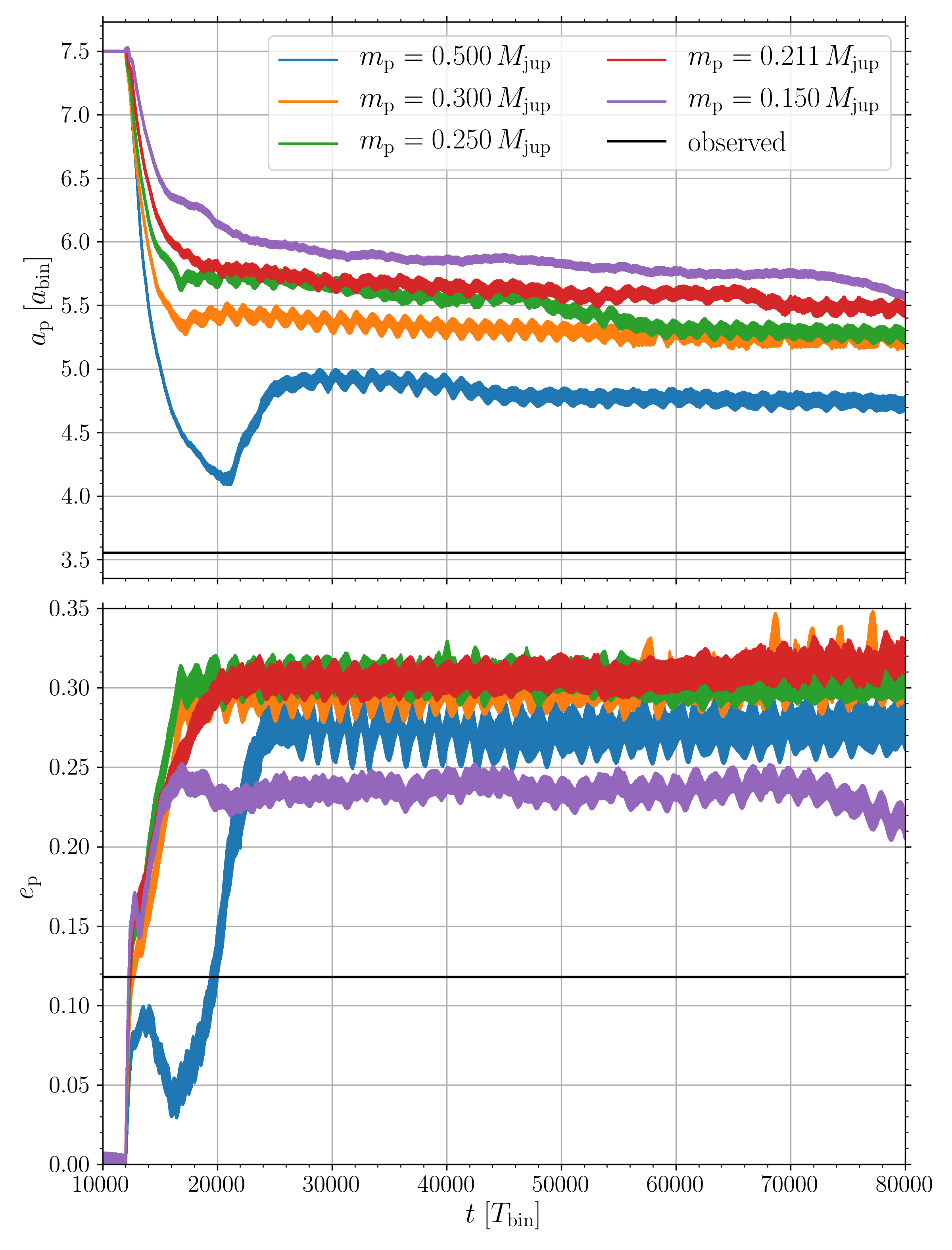}}
    \caption{Orbital elements of migrating planets in a circumbinary disc around
    the Kepler-413 system with different masses. \emph{Top:} Semi-major axis of
    the planet. \emph{Bottom} Eccentricity of the planet. The black lines show
    the observed values. The observed mass of Kepler-413b is 
    $m_\mathrm{p} = 0.211\,M_\mathrm{jup}$.}
    \label{img:k413_orbital_elements}
\end{figure}
\subsection{Kepler-413}
For Kepler-413, we model embedded planets with masses ranging from $m_\mathrm{p} =
0.150\,M_\mathrm{jup}$ up to $m_\mathrm{p} = 0.500\,M_\mathrm{jup}$; the observed
mass is $m_\mathrm{p} = 0.211\,M_\mathrm{jup}$.  Concerning the disc dynamics, Kepler-413
is very similar to Kepler-34; they both show a gap with high eccentricity. As a
consequence no low-eccentricity state seems to exist for the planet as shown in
Fig.~\ref{img:k413_orbital_elements}.  Again, all planets show more or less the
same migration behaviour: a fast inward migration, followed by a short period of
outward migration (only visible for massive planets), and then a fast arrival at
their final position.  In addition, all planet orbits are aligned with the inner
gap, as in Kepler-34.  The time of alignment depends on the planet mass and is
earlier for the lightest planets. The final orbital parameters of the simulated
planet with the observed mass can be found in Table~\ref{tab:orbital_elements}.

\section{Discussion}\label{sec:discussion}
Through 2D hydrodynamical simulations, we first studied the
structure of the circumbinary disc around five observed systems (Kepler-16, -34,
-35, -38, -413). After determining the structure of the inner disc (gap size and
eccentricity as well as precession period of the gap) we inserted planets of
various masses into our simulations and examined how they migrated through the disc
and how the presence of the planet changed the disc structure. In the following,
we summarise the most important results of our simulations with and without
planets.

\subsection{Disc dynamics}\label{ssec:sum_disc_dyn}
Firstly, we looked at the structure of the inner gap created by the gravitational
interaction between the binary and the disc. For simplicity, we restricted
ourselves here to locally isothermal discs with a fixed aspect ratio ($h=0.05$)
and viscosity ($\alpha=0.01$) and studied different binary parameters.  In
agreement with previous studies \citep{2017MNRAS.466.1170M, 2017A&A...604A.102T},
we found that for all systems, the circumbinary discs became eccentric and showed
a coherent slow prograde precession.  Plotting the obtained precession period
$T_\mathrm{prec}$ versus the obtained gap size $a_\mathrm{gap}$ , we confirmed the
existence of two different branches for varying binary eccentricities, as shown
in Fig.~\ref{img:Tprec_vs_agap}.  As seen in the figure, the four systems,
Kepler-16, 35, -38 and -413, fall exactly on the lower branch of the relation
found in our first study  \citep{2017A&A...604A.102T}. While Kepler-34 lays on
the extreme end of the upper branch, due to the very large binary mass ratio and
eccentricity of the system. To account for the
different binary parameters, the $y$-axis has been rescaled according to the
scaling of free particle orbits around binaries, eq.~\eqref{eq:particle_tprec}.

In addition to these two known branches, we discovered a new behaviour for small
binary eccentricities between $e_\mathrm{bin} = 0.0$ and $e_\mathrm{bin} =
0.05$.  Starting from circular binaries, the precession period of the gap as well
as the gap size increase up to $e_\mathrm{bin} = 0.05$. From there on, both
quantities decrease again, as expected on the lower branch.  At a critical
$e_\mathrm{bin} \approx 0.18,$ the trend reverses again. 
A more detailed investigation of this additional new loop in the
$T_\mathrm{prec}$-$a_\mathrm{gap}$ diagram lies beyond the scope of this paper
and is deferred to a future study.

In general, the five simulated systems behaved as expected and also the results
from the two different codes used in this study, namely \textsc{Pluto} and
\textsc{Fargo3D}, produced matching results.

\subsection{Evolution of embedded planets}\label{subsec:sum-planet-evol}
After we had investigated the disc structure without the presence of a
planet, we introduced planets to the simulations and studied how they migrated
through the discs. The migration of planets and especially the final orbital
parameters of the planets are important since it is assumed that planets in a
P-type orbit around a binary form in the outer part of the disc, where the
gravitational influence of the forming process due to the binary is negligible,
and migrate inward to their final observed position.

Firstly, the disc without a planet was simulated until a quasi-steady state was
reached, then we added the planet on a circular orbit for \num{2000} binary
orbits so that the disc could adjust to the planet. After that initialisation
period, the planet was allowed to freely move through the disc according to the
disc torques acting on it.  In the following, we summarise the main results using
the example of Kepler-38.
\begin{itemize}
    \itemsep1.0em
    \item \textbf{Initial planet position:} In a first simulation series, where
        we varied the initial planet position, we found that the final parking
        position of the planet does not depend on this initial planet location.
        In all cases, the planet migrates until it reaches the edge of the disc.
        At this point a strong positive corotation torque balances the negative
        Lindblad torque, responsible for inward migration, and the planet stops
        \citep{2006ApJ...642..478M}. 

    \item \textbf{Variation of planet mass:} When we varied the planet mass, we
        observed two different final configurations. In one state, the
        eccentricity of the planet is not excited and the planets remain on a
        nearly circular orbit around the binary. This low-eccentricity state
        applies for \enquote{massive} planets. These massive planets migrate
        smoothly through the disc until they reach their final position. During
        this migration process, they are able to alter the disc structure by
        shrinking and circularising the inner cavity of the disc. In contrast,
        the eccentricity of \enquote{lighter} planets becomes excited during their
        inward migration through the disc. Their migration path is not as smooth
        as in the massive planet case.  An initial period of rapid inward migration
        is typically followed by a short and fast period of outward migration,
        that is again followed by a period of inward migration but with a much
        slower speed than in the first phase. During this migration through the
        disc, the orbit of the planets becomes fully aligned with the precessing
        inner gap of the disc and they end up in a state of apsidal corotation.

        We were not able to find a sharp boundary between the massive and the
        light planets. The two different codes \textsc{Pluto} and
        \textsc{Fargo3D} agreed well on the general migration behaviour of the
        planets (two eccentricity states) but showed small differences
        concerning the boundary between the two states of massive and light
        planets.

    \item \textbf{Variation of disc mass:} A planet's behaviour is not only
        determined by its mass but also the planet-to-disc mass ratio. In a
        simulation with a massive planet and a disc with reduced mass we could
        observe that the massive planet behaves now like the light planets
        before, showing eccentricity excitation and orbital alignment. In other
        cases, we were able to switch the behaviour of a light planet to that of
        a massive one by decreasing the disc's mass.  Therefore, whether a planet in
        its final starting position shows an orbital alignment with the disc and
        an eccentric orbit or a more circular non-aligned orbit depends on the
        mass ratio between disc and planet.
    
    \item \textbf{Resonances:} During our simulations, we quite often observed
        that planets were captured into a high-order resonance (11:1, 9:1, 8:1)
        with the binary. Planets which are trapped in a resonance do not migrate
        further to their observed location and their eccentricity is excited.
        We observed the capture into a resonance only for the \enquote{light}
        planet case during the second, slower inward migration phase. 

\end{itemize}

The existence of two different eccentricity states was observed for the three
systems Kepler-16, -35, and -38. For the systems Kepler-34 and Kepler-413, we could
only observe the high eccentricity state for the light planet case with
full orbital alignment between disc and planet.  Even for planets with a mass
around half a Jupiter mass or more, we were not able to trigger the massive planet
case for these two systems. The reason for this behaviour could be the initial
large and eccentric disc gaps in Kepler-34 and Kepler-413. Although the heavier
planets were able to partly circularise the inner gap, they could never decrease
the discs eccentricity below a value of \num{0.1}.  The two eccentricity states
depend on the planet-to-disc mass ratio as well as on the binary parameters,
since the latter determine the size and eccentricity of the disc's gap for
otherwise identical disc parameter like pressure and viscosity. 

In this paper, we did not study the influence of the disc's pressure and
viscosity.  In all simulations, we chose an aspect ratio of $h=0.05$ and a
viscosity of $\alpha = 0.01$ to be consistent with our first paper, although an
aspect ratio of $h=0.05$ leads to rather high gas temperatures at the location
of the planet. A comparison with a test simulation that also considered viscous
heating and radiative cooling (for the same disc mass) suggests that an aspect
ratio of $h=0.04$ represents a more realistic gas temperature.

Test simulations of circumbinary discs around Kepler-38 with $\alpha
=0.001$ and $\alpha = 0.05$ show that in the case of low viscosity the observed
planet with mass $m_\mathrm{p} = 0.384\,M_\mathrm{jup}$ behaves accordingly to
the light case, whereas in the case of high viscosity all three simulated planets
with masses $m_\mathrm{p} = \{0.384,\ 0.300,\ 0.250\}\,M_\mathrm{jup}$ are able
to dominate the disc. Since higher viscosity leads to a higher mass flow through
the disc, the disc with $\alpha=0.05$ is less massive than in our standard case;
therefore all simulated planets are able to shrink and circularise the gap. In
the case of $\alpha = 0.001,$ the disc is more massive, meaning that even the planet
with mass $m_\mathrm{p} = 0.384\,M_\mathrm{jup}$ behaves like a light planet.

In general, we see both migration regimes if we vary the viscosity or the
aspect ratio of the disc; only the final orbital parameters can change.
This influence of the disc's pressure and viscosity on the final orbital
parameters of the planet needs to be examined in future studies together with a
more realistic equation of state rather than the isothermal one used in this study.

\section{Summary}\label{sec:summary}
In all our simulations, the planet migration stopped at the inner boundary of the
disc.  However, the general observation in all five Kepler systems considered,
was that the planets stop their migration outside of the observed location; see
Table~\ref{tab:orbital_elements}.  Two of the modelled systems  (Kepler-16 and
-38) ended up in the non-eccentric state where the final stopping point was
about 30\% larger than the observed locations, while the eccentricities were more
or less comparable to the observed values.  Kepler-35b reached an intermediate
case where the final eccentricity was $\ep = 0.12$ and a semi-major axis which
was 30\% larger than the observed one. Two
systems (Kepler-34 and -413) ended up in the high eccentric, aligned state with a
calculated final position about 1.5 times larger than the observed location and
calculated eccentricities about 2-2.5 times larger than the observed ones.
Although the final orbital elements did not match the observed ones very well,
our simulations nevertheless did reproduce the fact that the systems with the
highest gap eccentricity (Kepler-34 and -413) have the planets with the highest
observed eccentricity among all systems considered in this study. We therefore
suggest that the large observed planet eccentricities for these systems are
caused by an eccentric circumbinary disc.

Nevertheless, because the calculated final position of the planets is too large
in all cases, a mechanism is needed that is able reduce the inner gap size of
the disc to allow the planets to migrate further into the observed
locations. One such mechanism could be disc self-gravity as described
in~\citet{2017MNRAS.469.4504M}, although, according to the authors, only for very
massive discs with masses ten times the minimum mass solar nebula or more, will the
inner gap shrink further. Another way to reduce the disc's gap size could be
to change the disc parameters, such as pressure and viscosity. Increasing the
temperature and/or the viscosity will produce smaller gaps.  But as calculations
in our first study~\citep{2017A&A...604A.102T} have shown, possibly unusually
high pressures or viscosities may be needed to reduce the gap size substantially. 

Furthermore, the impact that radiative effects have on the
inner gap of the disc should be investigated. A good starting point will be simulations that include
viscous heating and radiative cooling from the disc surface. Even more
sophisticated simulations could also consider the radiative transport in the
disc's midplane and irradiation from the central binary.  Finally, 
 whether full 3D simulations will show different results remains to be seen.  Considering
the long timescales to be simulated, this is still numerically very challenging.

\begin{acknowledgements}
Daniel Thun was funded by grant KL 650/26 of the German Research Foundation
(DFG).  Most of the numerical simulations were performed on the bwForCluster
BinAC. The authors acknowledge support by the High Performance and
Cloud Computing Group at the Zentrum für Datenverarbeitung of the University
of Tübingen, the state of Baden-Württemberg through bwHPC and the German
Research Foundation (DFG) through grant no INST 37/935-1 FUGG. All plots in
this paper were made with the Python library matplotlib \citep{Hunter:2007}.
\end{acknowledgements}

\bibliography{cb_planets}
\bibliographystyle{aa}

\begin{appendix}
\section{N-body solver for \textsc{Pluto}}\label{sec:pluto_nbody}
The N-body solver for \textsc{Pluto} was developed by ourselves in the framework
of this project and we briefly discuss some implementation details. In the
discussion, we restrict ourselves to the 2D polar case. The
\textsc{Pluto}-code solves the hydrodynamic equations with the help of a
finite-volume method which evolves volume averages in time. For the time
evolution, we use the second-order Runge-Kutta method (RK2) provided by
\textsc{Pluto}. Given the hydrodynamical variables $\vec{V}^n = (\varrho^n,
u_R^n, u_\varphi^n)$ at time $t^n$, the positions $\vec{x}_\ell^n$ and
velocities $\vec{v}_\ell^n$ of the $N$ bodies ($\ell = 1, ... N$) also at
time $t^n$, the whole system (hydrodynamical variables and $N$ bodies) are
advanced from time $t^n$ to $t^{n+1} = t^n + \Delta t$ using the following RK2 method:
\begin{align}
    \vec{U}^* &= \vec{U}^n + \Delta t \mathcal{L}^n \\
    \vec{U}^{n+1} &= \frac{1}{2}
                     \left(\vec{U}^n + \vec{U}^* + \Delta t \mathcal{L}^*\right) \,.
\end{align}
For a definition of the \enquote{right-hand-side} operator $\mathcal{L}$ see
\citet{2007ApJS..170..228M}. Here, $\vec{U} = (\varrho, \varrho u_R, \varrho
u_\varphi)$ denotes the vector of conserved variables. The intermediate
predictor state, $\vec{U}^*$, is given at time $t^{n+1}$ which is important
for the correct stepping of the $N$ bodies. During the first RK2-substep
(called A) we perform the following substeps:
\begin{itemize}
    \item [A1] Set boundaries.
    \item [A2] Calculate disc feedback on each body $\ell$ (second term in
        eq.~\eqref{eq:eq_of_motion}) 
        \begin{equation}
            \begin{split}
            \vec{a}^n_{\mathrm{disc},\ell} \left(\vec{U}^n\right) 
                = &\sum_{ij} 
                   \frac{G \Sigma_{ij}^n \dd V_{ij}}
                        {\left[ (x_{c,ij}-x_\ell^n)^2+(y_{c,ij}-y_\ell^n)^2 
                               +(\varepsilon h R_i)^2\right]^{3/2}}\\
                  &\qquad 
                   \times\begin{pmatrix}x_{c,ij} - x_\ell^n\\
                                       y_{c,ij} - y_\ell^n \end{pmatrix}
            \end{split}
        \end{equation}
        where $x_{c,ij} = R_i \cos(\varphi_j)$ and $y_{c,ij} = R_i
        \sin(\varphi_j)$ denote the $x$- and $y$-coordinate of grid cell $(i,j)$. 
      The volume element $\dd V_{ij}$ of this cell is given in polar coordinates by 
        \begin{equation}
            \dd V_{ij} = \frac{1}{2}
                         \left(R_{i+\frac{1}{2}}^2 - R_{i-\frac{1}{2}}^2 \right) 
                         \times\left( \varphi_{j+\frac{1}{2}} 
                                    -\varphi_{j-\frac{1}{2}}\right)
        .\end{equation}
    \item [A3] Calculate the acceleration (disc force and mutual gravitational attraction)
         on each of the $N$ bodies (eq.~\eqref{eq:eq_of_motion}, first two terms)
        \begin{align}
            \label{eq:accel}
             \vec{a}_\ell^n =& \vec{a}^n_{\mathrm{grav},\ell}
                              \left(\vec{x}^n_1,\dotsc,\vec{x}^n_N\right) 
                             +\vec{a}^n_{\mathrm{disc},\ell}
                              \left(\vec{U}^n\right) \nonumber  \\ \nonumber
                           =& -\sum_{k \neq \ell}^{N} 
                               \frac{G M_k}
                                    {\left[ (x_\ell^n-x_k^n)^2 
                                           +(y_\ell^n-y_k^n)^2\right]^{3/2}} 
                               \begin{pmatrix}x_\ell^n-x_k^n \\ 
                                              y_\ell^n-y_k^n \end{pmatrix} \\  
                            &+\vec{a}^n_{\mathrm{disc},\ell} 
                              \left(\vec{U}^n \right.).
        \end{align}
    \item [A3] Calculate the indirect term, due to the acceleration of the
        centre of mass of the binary (eq.~\eqref{eq:eq_of_motion}, here 1 is the
        index of the primary and 2 is the index of the secondary).
        \begin{equation}
            \vec{a}^n_\mathrm{com} = \frac{M_1 \vec{a}^n_1 + M_2
            \vec{a}^n_2}{M_1 + M_2}
        .\end{equation}
    \item [A4] Hydro update $\vec{U}^{*} = \vec{U}^n + \Delta t  \mathcal{L}(\vec{U}^n)$.
        During this step we need to calculate the
        potential (eq.~\eqref{eq:potential}) at each cell location $(i,j)$
        \begin{align}
            \Phi_{ij}^{n} =& -\sum_{\ell=0}^N 
                              \frac{G M_\ell}
                                   {\sqrt{ (x_{c,ij}-x^n_\ell)^2 
                                          +(y_{c,ij}-y^n_\ell)^2 
                                          +(\varepsilon h R_i)^2}},\\
                           &+ \vec{a}_\mathrm{com}^n \cdot 
                              \begin{pmatrix}x_{c,ij} \\ y_{c,ij}\end{pmatrix}.
        \end{align}
\end{itemize}
During the second RK2-substep (B) the following steps are necessary:
\begin{itemize}
    \item [B1] Set boundaries.
    \item [B2] Advance bodies from $\vec{x}^n$ to $\vec{x}^{n+1}$ using a
        fourth order Runge-Kutta integrator (RK4) using the acceleration from eq.~(\ref{eq:accel}).
        In every substep of the RK4 integrator the mutual gravitational
        accelerations of the bodies are recalculated using the
        new substep positions (first term in equation~\eqref{eq:eq_of_motion}).
        For the disc feedback we use $\vec{a}_{\mathrm{disc},\ell}^n$ in every
        substep. After the $N$-body update we transform all quantities back to the
        centre of mass (let $\vec{X}^{n+1}$ and $\vec{V}^{n+1}$ denote the position and
        velocity of the centre of mass of the binary):
        \begin{align}
            \vec{x}^{n+1}_\ell &= \vec{x}^{n+1}_\ell - \vec{X}^{n+1} \\
            \vec{v}^{n+1}_\ell &= \vec{v}^{n+1}_\ell - \vec{V}^{n+1.}
        \end{align}
        This simple shift accounts for the indirect term $\vec{a}_\mathrm{com}$
        in eq.~\eqref{eq:eq_of_motion}.
    \item [B2] Calculate acceleration of each body
        (eq.~\eqref{eq:eq_of_motion}) using the updated body positions
        $\vec{x}^{n+1}$ and the updated hydro variables $U^{*}$.
    \item [B3] Calculate the indirect term, due to the acceleration of the
        centre of mass of the binary (eq.~\eqref{eq:eq_of_motion}) using the
        accelerations of the binary from step B2.
    \item [B4] Hydro update $\vec{U}^{n+1} = \frac{1}{2}\left(\vec{U}^n + \vec{U}^* +
        \Delta t \mathcal{L}(U^{*}) \right)$. During this step we need to calculate 
        the potential (eq.~\eqref{eq:potential}) at each cell location.
\end{itemize}
Now all quantities (hydro and $N$-body positions) are at time $t^{n+1}$ and a new
RK2 cycle starts.

\section{Comparison with \textsc{Fargo3D}}\label{sec:comparison_with_fargo}
To test the implementation of our $N$-body solver for \textsc{Pluto} we compared
circumbinary disc simulations with an embedded planet for the Kepler-38 system
with \textsc{Fargo3D} simulations (used in a 2D setup). 
\begin{figure}
    \centering
    \resizebox{\hsize}{!}{\includegraphics{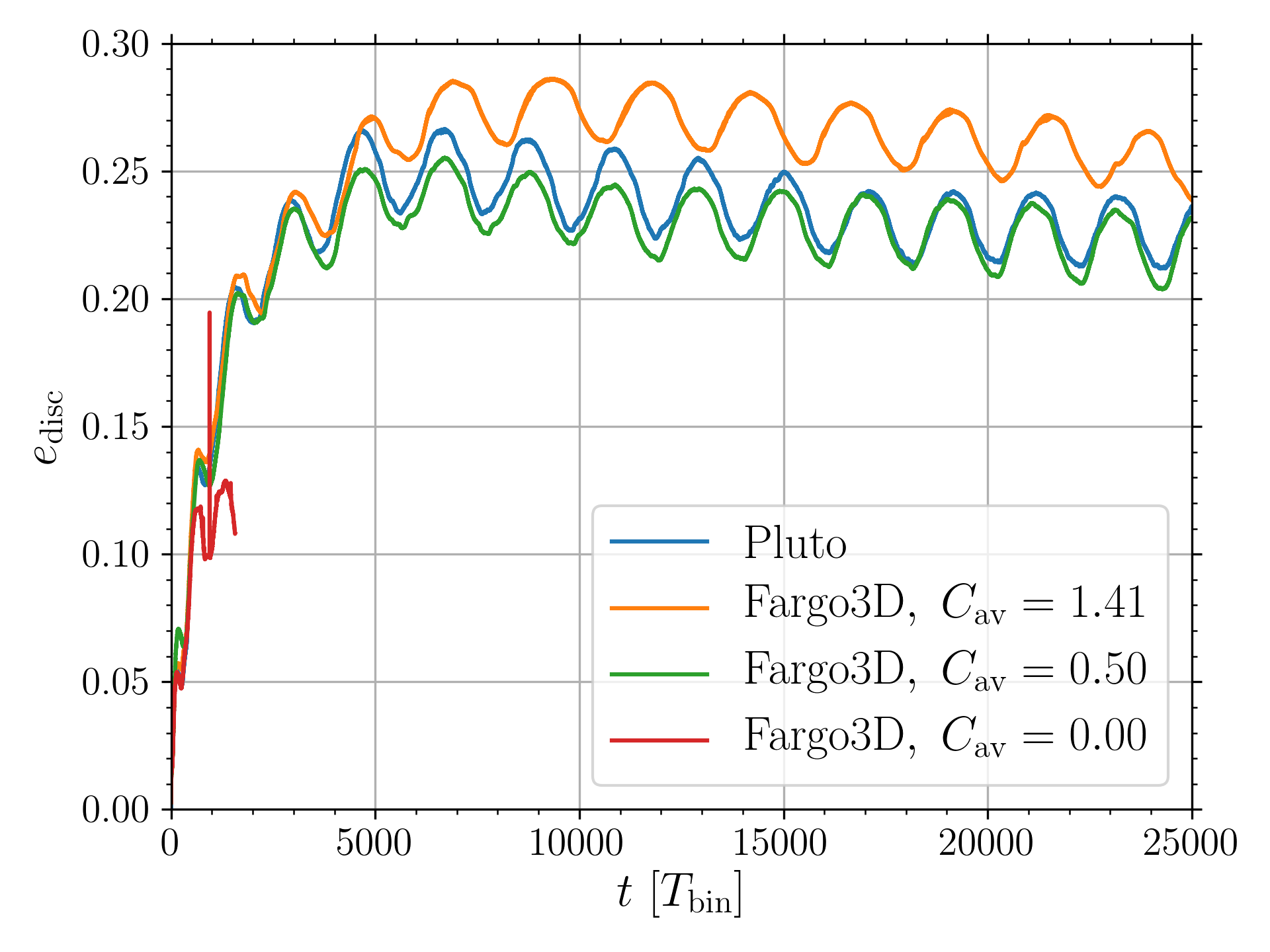}}
    \caption{Time evolution of the disc eccentricity for the Kepler-38 system.
    The blue curve shows a \textsc{Pluto} simulation, while all other curves
    show results from \textsc{Fargo3D} simulations, where the strength of the
    artificial viscosity was varied.}
    \label{img:k38_low_av}
 \end{figure}
\begin{figure*}
    \centering
    \includegraphics[width=17cm]{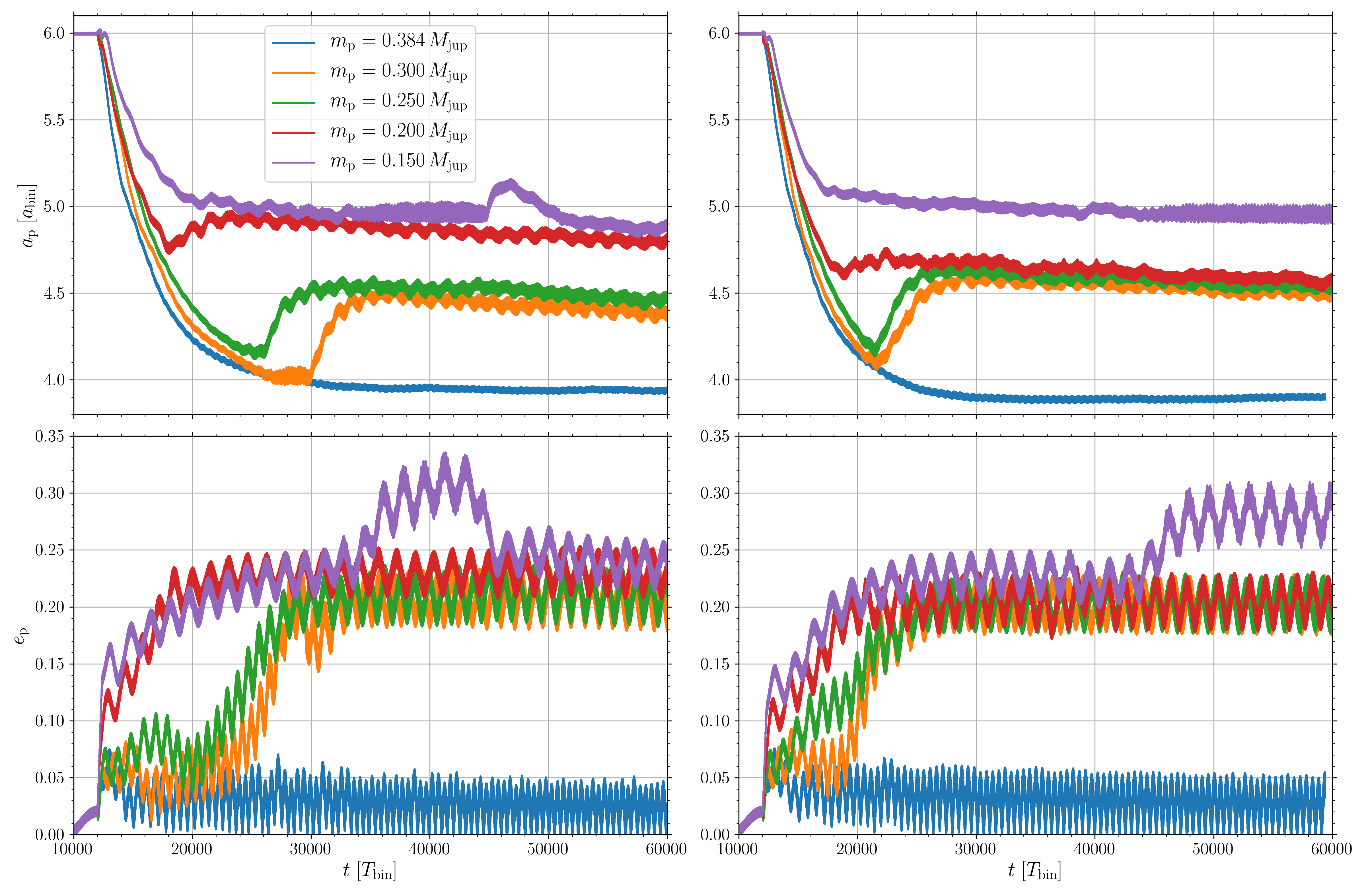}
    \caption{Orbital elements of planets with different mass. \emph{Top}:
    Semi-major axis of the planets. \emph{Bottom}: Eccentricity. \emph{Left
    column}: \textsc{Pluto} simulations. \emph{Right column}: \textsc{Fargo3D}
    simulations.}
    \label{img:k38_cmp_orbital_elements}
\end{figure*}

As briefly discussed in Sect.~\ref{ssec:numerics}, \textsc{Fargo3D} uses an
artificial viscosity (AV) to smooth out shocks over various cells. The strength
of the artificial viscosity is controlled by the constant $C_\mathrm{av}$. This
constant is basically the number of cells over which the shock is smoothed.  To
study the impact of AV on the disc dynamics, we varied the constant
$C_\mathrm{av}$ leaving the physical setup unchanged Figure~\ref{img:k38_low_av}
shows that the strength of the AV can change the simulation outcome
significantly.  Shown is the time evolution of the inner disc eccentricity for
the Kepler-38 system.  The eccentricity obtained with the standard artificial
viscosity constant of \textsc{Fargo3D} ($C_\mathrm{av}=1.41$, orange curve) does
not reproduce the \textsc{Pluto} result (blue curve). To obtain a better
agreement between the codes, we had to reduce the artificial viscosity used in
\textsc{Fargo3D} to $C_\mathrm{av} = 0.5$.  Using these adjustments, the new
\textsc{Fargo3D} results (green curve in Fig.~\ref{img:k38_low_av}) and the
\textsc{Pluto} agree very well indeed.  Simulations with no artificial viscosity
are not stable at all (red curve in Fig.~\ref{img:k38_low_av}).

For the rest of this section, we compare \textsc{Pluto} and \textsc{Fargo3D}
simulations of a circumbinary disc around Kepler-38 with an embedded planet.
Figure~\ref{img:k38_cmp_orbital_elements}
shows the semi-major axis (top row) and the eccentricity time evolution
(bottom row) of five planets with different masses from
$m_\mathrm{p}=0.150\,M_\mathrm{jup}$ to $m_\mathrm{p}=0.384\,M_\mathrm{jup}$. The left column
shows \textsc{Pluto} results whereas the right column shows \textsc{Fargo3D}
results.
The codes agree very well for the $m_\mathrm{p} = 0.384\,M_\mathrm{jup}$ case (blue curve
in Fig~\ref{img:k38_cmp_orbital_elements}). In both
simulations the planet smoothly migrates inward until it reaches a final position
slightly below $4\,a_\mathrm{bin}$. In the \textsc{Fargo3D} simulations the
planet migrates faster and a little bit more inward. Additionally, in both
simulations the eccentricity of the massive planet stays below 0.05. 

The codes produce also in the $m_\mathrm{p} = 0.150\,M_\mathrm{jup}$ case the same
results. This lighter planet does not migrate as far inward as the massive planet
discussed earlier. It reaches its final parking position at roughly
$5\,a_\mathrm{bin}$, but its eccentricity is increased to a value between 0.2
and 0.25. In both simulations the planet is captured in a 11:1 resonance with
the binary. The only difference here is that in the \textsc{Pluto} simulation
the planet is kicked out of this resonance while in the \textsc{Fargo3D}
simulations the planet remains in resonance with the binary. A more detailed
analysis of the resonance can be found in Sect.~\ref{sssec:var_planet_mass}.

For the planets with masses $m_\mathrm{p} = 0.250\,M_\mathrm{jup}$ and $m_\mathrm{p} =
0.300\,M_\mathrm{jup}$ , the two codes produce the same overall behaviour but
deviate in the details. As discussed in Sect.~\ref{sssec:var_planet_mass}, these
planets migrate first inward, then they reverse their migration direction and
undergo a period of fast outward migration. After that they start to migrate
inward again but on a much longer timescale than during the first inward
migration period. This behaviour is observed with both codes. In the
\textsc{Fargo3D} simulations the planets start their outward migration period in
both cases a bit earlier than in the \textsc{Pluto} simulations. A reason why in
the \textsc{Pluto} simulations the planets' outward migration period starts at
different times could be the fact that the $m_\mathrm{p} = 0.300\,M_\mathrm{jup}$ planet
is briefly captured in a 8:1 resonance with the binary between
$t=\num{27000}\,T_\mathrm{bin}$ and $t=\num{30000}\,T_\mathrm{bin}$. But besides
these differences the codes produce the results in good agreement for the final
orbital parameters of the planets. A final position around $4.5\,a_\mathrm{bin}$
and an increased eccentricity of around 0.20.

For the case $m_\mathrm{p} = 0.200\,M_\mathrm{jup}$ , the codes differ in the final orbital
elements of the planet. In the \textsc{Pluto} simulations, this planet ends up
with orbital parameters like the light planet with mass $m_\mathrm{p} =
0.150\,M_\mathrm{jup}$, whereas in the \textsc{Fargo3D} simulations the planet's
final orbital parameter are similar to the intermediate planets with
$m_\mathrm{p}=0.250\,M_\mathrm{jup}$ and $m_\mathrm{p}=0.200\,M_\mathrm{jup}$.  As discussed in
the main part of this paper, we observe a different migration behaviour and
different final orbital elements for light and massive planets. For very massive and
very light planets the codes produce consistent and matching results.  However,
the exact point (in mass) at which the transition between massive and lighter
planets occurs depends somewhat on the code used.  Whether intermediate mass
planets land on the massive or light branch depends on the code, but the absolute
difference in mass is not very large.  This was not only observed for the
Kepler-38 system but also for Kepler-16 and Kepler-35.

For example, a \textsc{Fargo3D} simulation of Kepler-35 with a planet of
mass $m_\mathrm{p} = 0.550\,M_\mathrm{jup}$ shows the same results as the \textsc{Pluto}
simulation.  But results of a \textsc{Fargo3D} simulation with planet mass
$m_\mathrm{p} = 0.500\,M_\mathrm{jup}$ shows the behaviour of a light planet
(eccentricity excitation, orbital alignment), whereas in \textsc{Pluto}
simulations the $m_\mathrm{p} = 0.500\,M_\mathrm{jup}$ planet is on the massive branch (see
Sect.~\ref{ssec:kepler35}).

For the Kepler-34 and -413 systems, we only observed one state with both codes.
In both codes, the large and highly eccentric inner gap always dominated the
planet.

Finally, we comment on the code performance on this problem.
In Table~\ref{tab:runtime}, we compare the average CPU-time per time step in seconds for
\textsc{Pluto} and \textsc{Fargo3D} for simulations of a disc around Kepler-38
with and without an embedded planet. The simulations that used a numerical resolution of 
 $684 \times 584$ grid cells were carried out on one or two Nvidia Titan X GPUs.
As seen, the two codes show a relatively similar performance in spite of the different numerical methods used.
For both codes, the runs on two GPUs are about 1.6 times faster that on a single GPU.
\begin{table}
    \caption{Average CPU-time per time step in seconds for \textsc{Pluto} and
    \textsc{Fargo3D} test simulations (using  $684 \times 584$ grid cells)
    with and without planet. The
    Fargo algorithm was not used in these simulations.}
    \label{tab:runtime}
    \centering
    \begin{tabular}{lcc}
        \hline\hline
        \noalign{\smallskip}
        Disc only &  \\
                       & 1 GPU & 2 GPUs \\
        \noalign{\smallskip}
        \hline
        \noalign{\smallskip}
        \textsc{Pluto}   & \num{1.00e-2} & \num{6.11e-3} \\
        \textsc{Fargo3D} & \num{9.52e-3} & \num{5.92e-3} \\
        \\
        \\
        \noalign{\smallskip}
        \hline
        \noalign{\smallskip}
        Disc with embedded &  &  \\
        migrating planet   & 1 GPU & 2 GPUs \\
        \noalign{\smallskip}
        \hline
        \noalign{\smallskip}
        \textsc{Pluto}   & \num{1.21e-2} & \num{7.47e-3} \\
        \textsc{Fargo3D} & \num{1.39e-2} & \num{8.30e-3} \\
        \noalign{\smallskip}
        \hline
    \end{tabular}
\end{table}
\balance

\end{appendix}

\end{document}